\documentclass[a4paper,11pt]{article}
\usepackage{jcappub} 
\usepackage{lineno}

\usepackage{lipsum}
\usepackage{url}

\usepackage[T1]{fontenc}
\usepackage{subcaption}
\usepackage{xspace}
\usepackage{booktabs}
\usepackage{multirow}
\usepackage{rotating}
\usepackage{csquotes}
\usepackage{siunitx}
\DeclareSIUnit\parsec{pc}
\DeclareSIUnit \parsec{pc}
\DeclareSIUnit \year{yr}
\DeclareSIUnit \steradian{sr}
\DeclareSIUnit \solarmass{\text{\ensuremath{{M}_{\odot}}}}
\DeclareSIUnit \solarluminosity{\text{\ensuremath{{L}_{\odot}}}}
\DeclareSIUnit \h{\ensuremath{\mathnormal{h}}}
\DeclareSIUnit \hubble{\h}
\DeclareSIUnit \hseventy{\ensuremath{\mathnormal{h_{70}}}}
\DeclareSIUnit \hMpc{\per\h\mega\parsec}
\DeclareSIUnit \micron{\micro\meter}
\DeclareSIUnit \kms{\kilo\meter\per\second}
\DeclareSIUnit \kmsMpc{\kilo\meter\per\second\per\mega\parsec}
\DeclareSIUnit{\squaredegree}{deg^2}

\newcommand{\ufig}{\texttt{UFig}\xspace}
\newcommand{\galsbi}{\texttt{galsbi}\xspace}
\newcommand{\sextractor}{\texttt{SExtractor}\xspace}
\newcommand{\pinocchio}{\texttt{PINOCCHIO}\xspace}
\newcommand{\mlim}{M_\mathrm{lim}}
\newcommand{\tq}{t_\mathrm{quench}}
\newcommand{\sigmlim}{\sigma_{M_\mathrm{lim}}}
\newcommand{\sigtq}{\sigma_{t_\mathrm{quench}}}
\newcommand{\fq}{f_\mathrm{0,quench}}

\defcitealias{fischbacherGalSBIPhenomenologicalGalaxy2025}{F25G}
\defcitealias{fischbacherSHAMOTRapidSubhalo2025}{F25S}
\defcitealias{bernerFastForwardModelling2024}{B24}

\title{
GalSBI: Forward Modelling Galaxy Clustering and Population
}

\author[a,1]{Silvan~Fischbacher,} 
\note[1]{Corresponding author.}
\author[b]{Luca~Tortorelli,}
\author[c]{Tomasz~Kacprzak,}
\author[a]{Alexandre~Refregier}

\vspace{0.2cm}

\affiliation[a]{Institute for Particle Physics and Astrophysics, ETH Zurich, Wolfgang-Pauli-Strasse 27, CH-8093 Zurich, Switzerland}
\affiliation[b]{University Observatory, Faculty of Physics, Ludwig-Maximilian-Universit{\"a}t M{\"u}nchen,
Scheinerstrasse 1, 81679 Munich, Germany}
\affiliation[c]{University of Applied Sciences Northwestern Switzerland FHNW, Bahnhofstrasse 6, CH-5210 Windisch, Switzerland}

\emailAdd{silvanf@phys.ethz.ch}

\abstract{
Forward modelling is a powerful approach for analyzing large-scale structure surveys.
For this purpose, we extend the GalSBI framework to jointly model the galaxy population and clustering using an efficient subhalo abundance matching scheme based on optimal transport.
We use simulation-based inference to constrain the model parameters by comparing \ufig image simulations with DES Y3 imaging data.
As a validation, we find that galaxy photometry and morphology agree well with multi-band imaging data of different depths, namely DES and HSC deep fields.
Galaxy clustering for simulation and data is also in good agreement when comparing the angular power spectrum for different magnitude and color cuts.
We further compare simulated redshift distributions against high-precision photometric redshifts in HSC deep field imaging of the COSMOS field.
We find the redshift distributions across magnitude cuts to be similar to previous work, however with more realistic uncertainty modelling due to the addition of clustering contribution to sample variance.
The agreement of the mean redshifts with data is very good, between $0.2\sigma$ and $1.6\sigma$ for different magnitude cuts, with sample variance being the dominant uncertainty contributor in bright samples ($<24$ mag) and subdominant compared to galaxy population model uncertainty in fainter samples.
As a byproduct we measure the galaxy luminosity function and galaxy-halo connection, which are broadly consistent with existing literature.
The updated GalSBI code and galaxy population model are publicly available.
They enable accurate forward-modelled image simulations with realistic clustering, which can be used to model the effect of sample variance, source clustering, redshift distributions, and blending in large-scale-structure surveys.
This makes GalSBI a powerful tool for the analysis of current and next-generation cosmological galaxy surveys.
}

\begin{document}
\maketitle
\flushbottom

\section{Introduction}
\label{sec:intro}
Since its first detection at the beginning of the century \citep{baconDetectionWeakGravitational2000,Kaiser:2000if,vanwaerbekeDetectionCorrelatedGalaxy2000,wittmanDetectionWeakGravitational2000}, weak gravitational lensing has been demonstrated to be a powerful cosmological probe.
Its constraining power can be further enhanced through the combination with galaxy clustering, typically referred to as $3\times2$pt analysis.
This combination further breaks parameter degeneracies and improves cosmological constraints thanks to their sensitivity to complementary systematic effects \citep{DESY6_3x2pt,heymansKiDS1000CosmologyMultiprobe2021,miyatakeHyperSuprimeCamYear2023,sugiyamaHyperSuprimeCamYear2023}.

Currently, Stage-III surveys such as the Dark Energy Survey \citep[DES,][]{darkenergysurveycollaborationDarkEnergySurvey2016}, the Kilo-Degree Survey \citep[KiDS,][]{dejongKiloDegreeSurvey2013}, and the Hyper Suprime-Cam Subaru Strategic Program \cite[HSC,][]{aiharaHyperSuprimeCamSSP2018} are publishing their final cosmological constraints while Stage-IV surveys, such as the Vera Rubin Observatory's Legacy Survey of Space and Time \citep[Rubin-LSST,][]{thelsstdarkenergysciencecollaborationLSSTDarkEnergy2021}, Euclid \citep{laureijsEuclidDefinitionStudy2011}, and the Nancy Grace Roman Space Telescope \citep[NGRST,][]{spergelWideFieldInfrarRedSurvey2015}, have started or will soon start to take their first data.
The increased statistical power of these surveys will place even more stringent requirements on systematic uncertainties, demanding improvements in both accuracy and precision.

An increasingly popular approach to achieve these improvements is forward modelling.
The core idea is to generate synthetic observations that are subject to the same systematic effects and selections as the real data.
This approach requires two main ingredients: a galaxy population model from which intrinsic properties can be sampled, and a model of the transfer function from intrinsic to observed properties.
Depending on the realism of these two parts, different applications can be unlocked.

One possible application is the estimation of photometric redshifts \citep[photo-$z$, see e.g.,][for reviews]{newmanPhotometricRedshiftsNextGeneration2022,tortorelliMachineLearningTechniques2026}.
This is one of the dominant sources of systematic uncertainties in current Stage-III surveys, where challenges with photo-$z$ estimation affected their final cosmological analyses, ranging from biases requiring mitigation \citep{dalalHyperSuprimeCamYear2023,liHyperSuprimeCamYear2023}, to the exclusion of redshift bins \citep{descollaborationDarkEnergySurvey2022,DESY6_3x2pt}, to shifts in cosmological constraints following improved calibration \citep{wrightKiDSLegacyRedshiftDistributions2025,wrightKiDSLegacyCosmologicalConstraints2025a}.
Forward modelling offers a complementary approach that avoids the need to correct for selection effects in spectroscopic calibration samples. 
By generating synthetic observations to which the same selection is applied as to the real data, it directly yields the ensemble redshift distribution without requiring a separate calibration step.
To do this, the galaxy population model and transfer function have to accurately model all galaxy properties relevant for the sample selection, at minimum the broadband magnitudes.

If the population model further includes morphology, the transfer function can be realistically modelled with image simulations, which additionally enables shear calibration \citep{brudererCosmicShearCalibration2018}.
Adding realistic galaxy clustering then unlocks a further set of applications: it allows quantification of sample variance when calibrating on small fields such as COSMOS, and combined with image simulations it improves blending calibration \cite[see][and references therein]{melchiorChallengeBlendingLarge2021} since blending is more likely in overdense regions.
Clustering of source galaxies is furthermore a systematic effect in weak lensing itself.
While it is negligible for standard 2-point analyses in Stage-III \citep{krauseDarkEnergySurvey2021a}, it already affects higher-order statistics \citep{gattiDetectionSignificantImpact2023a} and will become more important for Stage-IV.
Taken together, a forward model that jointly describes photometry, morphology and spatial positions opens the 
door to a fully self-consistent $3\times2$pt analysis.

Several frameworks have been developed along these lines with different modelling choices.
Our GalSBI model \citep[][hereafter \citetalias{fischbacherGalSBIPhenomenologicalGalaxy2025}]{fischbacherGalSBIPhenomenologicalGalaxy2025} \citep{tortorelliGALSBISPSStellarPopulation2025b} consists of a parametric description of the galaxy population and image simulations for the transfer function; it is described in detail below.
A similar parametric approach but with a data-driven noise model was used in \cite{alsingForwardModelingGalaxy2023,leistedtHierarchicalBayesianInference2023}.
Later work replaced the population model with a diffusion model \citep[\texttt{pop-cosmos,}][]{alsingPopcosmosComprehensivePicture2024} trained on COSMOS2020 catalog data \citep{weaverCOSMOS2020PanchromaticView2022}, further enabling individual photo-$z$ and physical property inference \citep{thorpPopcosmosScaleableInference2024,thorpPopcosmosInsightsGenerative2025a,degerPopcosmosStarFormation2025}.
The model was recently adapted to KiDS-1000 data \citep{leistedtPopcosmosForwardModeling2026,halderPopcosmosRedshiftsPhysical2026} calibrating their noise model on SKiLLS image simulations \citep{liKiDSLegacyCalibrationUnifying2023}, with an architecture consisting of a detection classifier and a conditional ODE flow, conceptually similar to the emulator developed in \citetalias{fischbacherGalSBIPhenomenologicalGalaxy2025}.

A complementary SPS-based forward model, PROVABGS, jointly fits photometry and spectra for the DESI Bright Galaxy Survey galaxies \citep{hahnDESIPRObabilisticValueAdded2023,hahnPROVABGSProbabilisticStellar2024}, providing full posterior distributions of physical properties for over 10~million galaxies via a neural-emulator-accelerated SPS model and enabling downstream hierarchical population inference \citep{liPopSEDPopulationLevelInference2024}.
A further forward modelling example is the Diffsky framework, built from fully differentiable models for stellar population synthesis \citep{hearinDSPSDifferentiableStellar2023}, halo mass assembly \citep{hearinDifferentiableModelAssembly2021}, galaxy assembly history \citep{alarconDiffstarFullyParametric2022}, and the star formation history \citep{alarconDiffstarPopGenerativePhysical2025}.

GalSBI builds on a series of work on forward modelling galaxy surveys, first proposed in \cite{refregierWayForwardCosmic2014}.
The galaxy population model uses parametric descriptions of observed statistical quantities such as the luminosity function, and the transfer function is applied using image simulations with the ultra-fast image generator \ufig \citep{bergeUltraFastImage2013,fischbacherUFigV1Ultrafast2025} or spectra simulations using \texttt{USpec} \citep{fagioliForwardModelingSpectroscopic2018,fagioliSpectroimagingForwardModel2020,lucatortorelliUSpec2UltrafastSpectrainprep.}.
The galaxy population model parametrization has been progressively improved \citep{herbelRedshiftDistributionCosmological2017,moserSimulationbasedInferenceDeep2024,fischbacherGalSBIPhenomenologicalGalaxy2025}, and applied to multiple surveys for improved constraints or validation (DES \citep{brudererCalibratedUltraFast2016,kacprzakMonteCarloControl2020}, CFHTLS \citep{tortorelliMeasurementBbandGalaxy2020}, PAUS \citep{tortorelliPAUSurveyForward2018,tortorelliPAUSurveyMeasurement2021}, HSC \citep{moserSimulationbasedInferenceDeep2024}).
The model, named GalSBI, was first released in \citetalias{fischbacherGalSBIPhenomenologicalGalaxy2025}, accompanied by an open-source Python package \cite{fischbacherGalsbiPythonPackage2025}.
Most recently, \cite{tortorelliGALSBISPSStellarPopulation2025b} introduced the GalSBI-SPS model, which replaces the spectral templates with a full stellar population synthesis framework using ProSpect \citep{robothamProSpectGeneratingSpectral2020}.
In all of this work, however, galaxy positions were assigned uniformly at random, neglecting the spatial clustering of galaxies.

In this work, we close this gap by extending GalSBI to model the galaxy-halo connection, obtaining realistic spatial distributions of galaxies in addition to their photometric and morphological properties.
This completes the set of ingredients needed for the applications outlined above, and in particular takes a significant step towards a fully forward-modelled $3\times2$pt analysis within the GalSBI framework.

\begin{figure}
    \centering
    \includegraphics[width=1\linewidth]{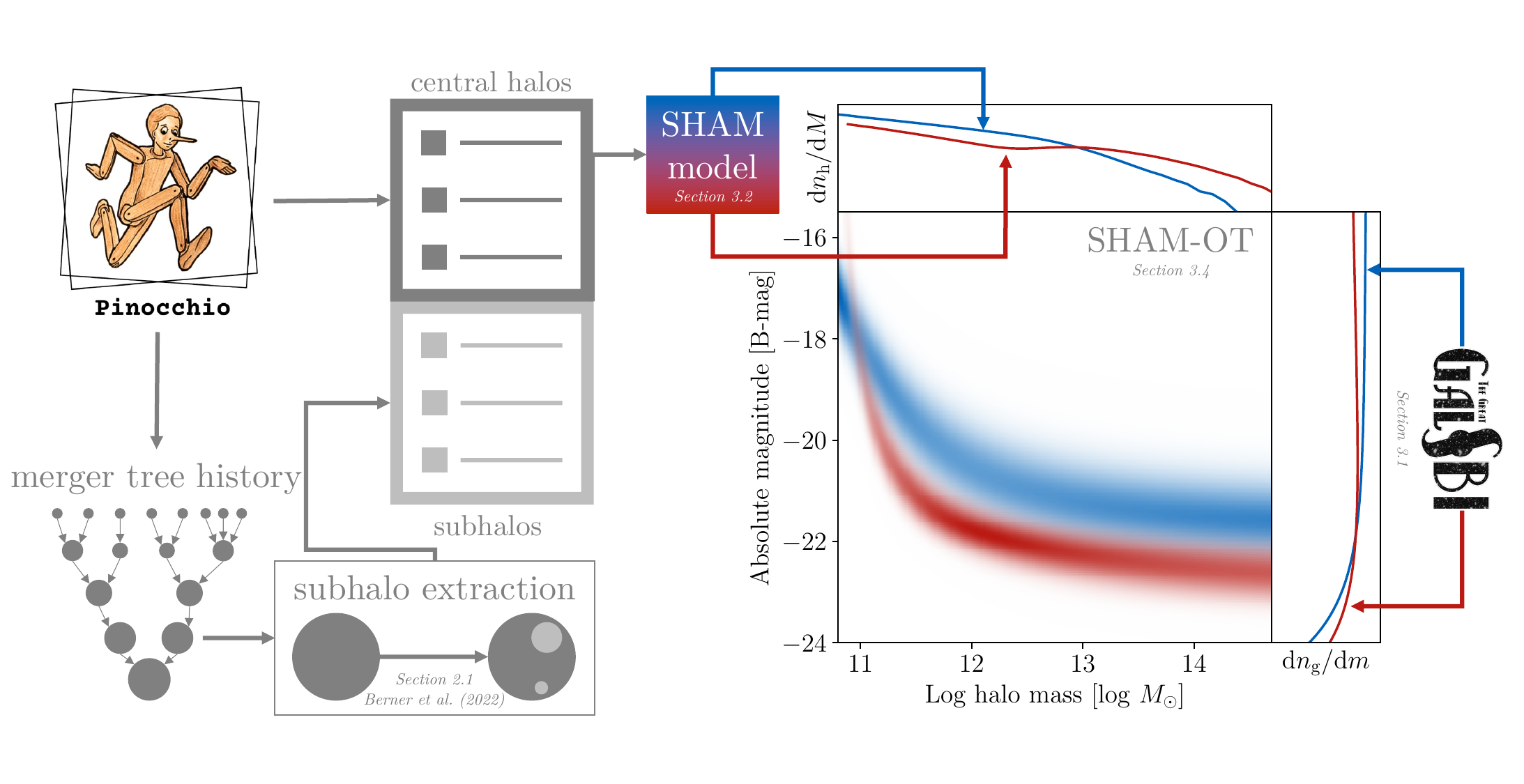}
    \caption{
    Illustration of the methodology used in this work.
    We use \pinocchio to create a halo catalog and the merger tree history.
    From the merger tree history, we extract subhalos modelling their survival time and sample their position within the parent halo from an isothermal profile (details are given in section \ref{sec:pinocchio}).
    The joint halo and subhalo catalog is then split into a red and blue halo mass function (details on the splitting are given in section \ref{sec:sham}).
    The galaxy population model (described in section \ref{sec:galpop}) defines analytical red and blue luminosity functions.
    These two normalized histograms are then matched using optimal transport to obtain the galaxy-halo connection (section \ref{sec:sham-ot}).
    }
    \label{fig:overview}
\end{figure}

To model the galaxy-halo connection, we use subhalo abundance matching \citep[SHAM,][]{kravtsovDarkSideHalo2004,valeLinkingHaloMass2004,conroyModelingLuminositydependentGalaxy2006}, which provides a simple yet accurate framework for populating dark matter only simulations with galaxies.
Previously, \cite[][hereafter \citetalias{bernerFastForwardModelling2024}]{bernerFastForwardModelling2024} used SHAM with a precursor of the GalSBI model to construct simulated catalogs with realistic positions, finding good agreement in the two-point function when comparing to DES Y1 data, but the computational cost of SHAM prohibited its use within fast image simulation pipelines.
This bottleneck was resolved by SHAM-OT \citep[][hereafter \citetalias{fischbacherSHAMOTRapidSubhalo2025}]{fischbacherSHAMOTRapidSubhalo2025}, which reformulates SHAM as an optimal transport problem.
Instead of sampling and sorting large galaxy and halo catalogs, the galaxy-halo connection is obtained directly by matching the halo mass function with the luminosity function using optimal transport, reducing the computational cost by orders of magnitude and making on-the-fly computation feasible.
Figure \ref{fig:overview} illustrates the methodology used in this work, combining methodological contributions from \citetalias{bernerFastForwardModelling2024}, \citetalias{fischbacherGalSBIPhenomenologicalGalaxy2025} and \citetalias{fischbacherSHAMOTRapidSubhalo2025}.

The paper is organized as follows.
We start in section~\ref{sec:sim} by presenting our halo and subhalo catalog simulation framework as well as the image simulation pipeline. 
In section~\ref{sec:methods}, we present the components of the extended GalSBI model and our inference approach.
We show our results in section~\ref{sec:results} before concluding in section~\ref{sec:conclusion}.
\section{Simulations and data}
\label{sec:sim}

In this section, we describe the simulation pipelines and data used in this work.
The methodologies for both the halo catalog simulation and the image simulations follow established approaches from prior work \citep{bernerRapidSimulationsHalo2022,kacprzakMonteCarloControl2020,moserSimulationbasedInferenceDeep2024} and are summarized here for completeness.

\subsection{Halo and subhalo catalog simulation}
\label{sec:pinocchio}
To assign realistic positions to simulated galaxies via subhalo abundance matching, we require a realistic halo and subhalo catalog.
The halos are simulated using \pinocchio \citep{taffoniPINOCCHIOHierarchicalBuildup2002,monacoPINOCCHIOPinpointingOrbitcrossing2002,monacoPredictingNumberSpatial2002,monacoAccurateToolFast2013,munariImprovingFastGeneration2017,rizzoSimulatingCosmologies$L$CDM2017}, which generates halo catalogs and merger histories by combining Lagrangian Perturbation Theory (LPT) with the Extended Press--Schechter formalism.
This allows \pinocchio to build groups of particles during the simulation, avoiding the need to track individual particle interactions at late times and resulting in a significant computational speedup compared to full N-body simulations.
The output is a halo lightcone catalog together with the full merger history.

Since \pinocchio does not directly simulate subhalo catalogs, we add subhalos using the methodology developed in \cite{bernerRapidSimulationsHalo2022} and further applied in \citetalias{bernerFastForwardModelling2024}.
Subhalos are extracted from the merger history by identifying progenitor groups that have merged into each host halo.
The mass of the subhalo is set to the mass of the progenitor at the time of merging, which is the only mass available from \pinocchio.
However, the mass at accretion is actually a better choice for subhalo abundance matching than the current subhalo mass, due to tidal stripping which affects the subhalo mass more strongly than the galaxy luminosity \citep{conroyModelingLuminositydependentGalaxy2006}.
It is therefore not a compromise arising from our use of \pinocchio, but a natural and preferred choice for subhalo abundance matching.

Two quantities must then be modelled: the survival time of each subhalo within its host, and its spatial position.
For the survival time, we use the fitting function presented in \cite{bernerRapidSimulationsHalo2022}, which depends on the host-to-subhalo mass ratio at accretion and includes a redshift dependence via the linear growth rate factor:
\begin{equation}
    \tau_\mathrm{merge} = A(D)\, \tau_\mathrm{dyn} \, \frac{(M_\mathrm{host}/m_\mathrm{sub})^{b(D)}}{\ln(1 + M_\mathrm{host}/m_\mathrm{sub})} \exp(c\,\eta) ,
\end{equation}
with the best-fit parameters
\begin{equation}
    A(D) = 0.195 \left[1 + \Theta(0.6 - D)\left(\left(\frac{D}{0.6}\right)^2 - 1\right)\right], \qquad b(D) = 0.92\,D(z), \qquad c = 1.9 ,
\end{equation}
where $\Theta$ is the Heaviside step function and $D(z)$ is the linear growth rate factor normalized to unity at $z=0$.
The remaining quantities are modelled rather than fitted: $\tau_\mathrm{dyn}$ is the dynamical time given by $0.1\,H(z)^{-1}$, and the orbital circularity $\eta$ is drawn from $P(\eta) \propto \eta^{1.2}(1-\eta)^{1.2}$ on the interval $[0.2,1]$.
A subhalo is considered to have survived if the time since accretion is shorter than $\tau_\mathrm{merge}$.

Surviving subhalos are then distributed within their host halos by drawing radial positions from an isothermal number density profile \citep{diemandVelocitySpatialBiases2004} with characteristic scale $0.37\,r_\mathrm{vir}$, and random angular positions.
This approach has been validated against N-body simulations using GADGET-2 \citep{springelGADGETCodeCollisionless2001,springelCosmologicalSimulationCode2005} with the halo finder ROCKSTAR \citep{behrooziRockstarPhaseSpaceTemporal2013}, showing good agreement in both the subhalo velocity function and the 2-point correlation function \citep{bernerRapidSimulationsHalo2022}.
The resulting halo and subhalo catalogs are then further rotated to maximize the number of independent realizations of the corresponding survey, for details see appendix~\ref{app:rotation}.

We run \pinocchio with a box size of $\SI{1000}{\hMpc}$, $4096^3$ particles, covering the redshift range $z \in [0, 5]$.
The minimum number of particles per halo is set to 10, resulting in a minimum (sub-)halo mass of $\SI{1.27e10}{\per\h\solarmass}$.
The cosmological parameters are set to the Planck~2018 fiducial values \citep{planckcollaborationPlanck2018Results2020}.

\subsection{Image simulations}
\label{sec:image_simulations}

\begin{figure}
    \centering
    \includegraphics[width=1\linewidth]{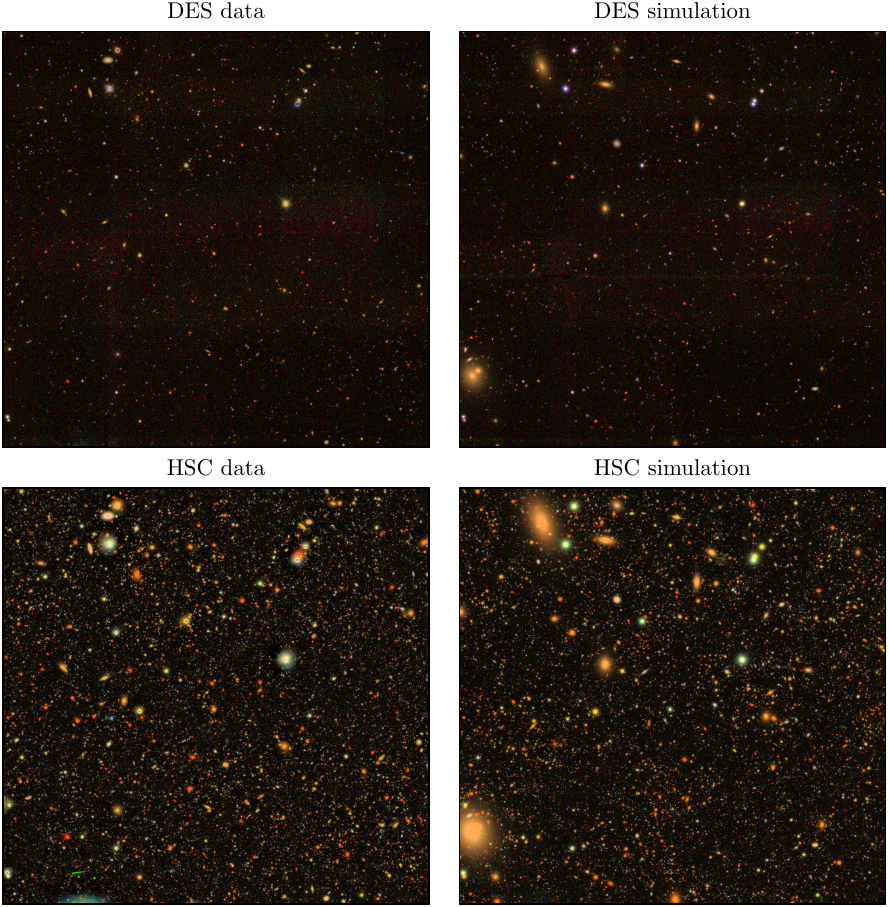}
    \caption{
    Real images (top) and corresponding simulation (bottom) of DES (left) and HSC (right) of the same patch of the sky.
    The RGB color image is generated using the method from \cite{luptonPreparingRedGreenBlueRGB2004} with $g,r,i$ bands.
    To match the coordinates of the two images, we show a zoomed version of the DES image.
    The galaxy population model corresponds to the fiducial model obtained in this work.
    }
    \label{fig:images}
\end{figure}

We perform image simulations with \ufig \cite{bergeUltraFastImage2013,fischbacherUFigV1Ultrafast2025}, examples of the image simulations used in this work are given in figure \ref{fig:images}.
For both surveys described below, we simulate the coadded images directly (rather than individual exposures), which is accurate for our purposes and substantially faster.
Stars are sampled from the Besançon model of the Milky Way \cite{robinSyntheticViewStructure2003}, accounting for stellar density variations across the sky, with the brightest stars position-matched to the Gaia catalog \cite{vallenariGaiaDataRelease2023}.
The point-spread function (PSF) is estimated at Gaia star positions in the real data using a convolutional neural network (CNN) approach \cite{herbelFastPointSpread2018,kacprzakMonteCarloControl2020}.
The predicted PSF parameters are then interpolated across the image using a Chebyshev polynomial basis (up to order 4), incorporating coadding information through the constructed exposure maps.

For details about the different configurations for source extraction and for the selection, we refer to Appendix \ref{app:selection}.

\subsubsection{Dark Energy Survey (DES)}
\label{sec:des_sims}

We use the publicly available Dark Energy Survey Year 3 data \citep[DES~Y3,][]{abbottDarkEnergySurvey2018a} covering roughly $5000\,\mathrm{deg}^2$, and follow the image simulation methodology of \cite{brudererCalibratedUltraFast2016,kacprzakMonteCarloControl2020}.
Background maps are created using the single-exposure information and CCD boundaries to model the sharp background variations in the coadds; see section III.C of \cite{kacprzakMonteCarloControl2020} for details.
The accuracy of this pipeline has been validated for cosmic shear measurements in \cite{kacprzakMonteCarloControl2020}, which imposes even stricter requirements on PSF and background than the present work.

We create two \sextractor \cite{bertinSExtractorSoftwareSource1996} catalogs from the real images.
The \emph{fiducial catalog}, used for validation, is detected on a $riz$ \texttt{CHI\_MEAN} image constructed following \cite{abbottDarkEnergySurvey2018a}.
During inference, we use an emulator (Section~\ref{sec:emulator}) that models the photometry in all bands without image simulations, except for the detection band(s).
To maximize its speed advantage, we simulate only the $i$-band for detection, and therefore construct the \emph{ABC-catalog} with $i$-band detection to ensure a consistent comparison between simulations and data.

\subsubsection{Hyper Suprime-Cam deep and ultra-deep fields (HSC DUD)}
\label{sec:hsc_sims}

We use the Hyper Suprime-Cam PDR3 deep and ultra-deep fields \citep[HSC~DUD][]{aiharaThirdDataRelease2022} and follow \cite{moserSimulationbasedInferenceDeep2024,fischbacherGalSBIPhenomenologicalGalaxy2025} for the \sextractor catalog construction and image simulation pipeline, with the $i$-band as detection image.
The main difference from the DES setup is the background estimation: rather than exposure-based maps, we use the \texttt{Background2D} function with the \texttt{SExtractorBackground} estimator and $3\sigma$ clipping from \texttt{photutils} \cite{bradleyAstropyPhotutils2302025}, which captures the smoother background variations present in HSC coadds.
For further details on the image simulation pipeline and quality assessment, we refer the reader to \cite{moserSimulationbasedInferenceDeep2024}.

The HSC deep fields partially overlap with the COSMOS2020 panchromatic photometric catalog \cite{weaverCOSMOS2020PanchromaticView2022}, which provides high-quality photometric redshifts via LePhare \cite{arnoutsMeasuringModellingRedshift1999,ilbertAccuratePhotometricRedshifts2006} and EAZY \cite{brammerEAZYFastPublic2008}.
This allows us to assign redshifts to galaxies in these overlapping regions, which will later be used during validation.
\section{Methods}
\label{sec:methods}

Building on the simulation infrastructure described in section \ref{sec:sim}, we extend the GalSBI model to include realistic galaxy clustering by introducing a model of the galaxy–halo connection and integrating it into the simulation-based inference (SBI) framework.
Photometry and morphology of simulated galaxies are modelled using the same methodology as in \citetalias{fischbacherGalSBIPhenomenologicalGalaxy2025} and is further described in section \ref{sec:galpop}.
Realistic positions are assigned using subhalo abundance matching (SHAM) in the efficient implementation presented in \citetalias{fischbacherSHAMOTRapidSubhalo2025}.
We give an overview of SHAM in section \ref{sec:sham}, an introduction to optimal transport (OT) in section \ref{sec:ot} and describe how OT can be used for SHAM in section \ref{sec:sham-ot}.
Section \ref{sec:cl} shows how we compute clustering statistics on the galaxy catalogs and section \ref{sec:sbi} presents our SBI pipeline.

\subsection{Galaxy population model}
\label{sec:galpop}
GalSBI is a parametric galaxy population model with two subpopulations: red (quiescent) and blue (star-forming) galaxies with mostly identical parametrizations but independent parameters.
We use the same parametrizations as the fiducial one from \citetalias{fischbacherGalSBIPhenomenologicalGalaxy2025}.
Galaxy absolute magnitudes and redshifts are drawn from luminosity functions.
Each galaxy is assigned a spectral energy distribution (SED) through a linear combination of the \texttt{kcorrect} templates \citep{blantonKcorrectionsFilterTransformations2007}.
After accounting for galactic extinction, the apparent magnitude is computed by integrating the SED over the relevant filter band.

Galaxy sizes are sampled from a lognormal distribution whose mean and scatter depend on absolute magnitude and redshift, capturing the observed size-luminosity relation.
The morphological parameters (absolute ellipticity and Sérsic index) are described by flexible Beta and Betaprime distributions respectively.
For further details, we refer to \citetalias{fischbacherGalSBIPhenomenologicalGalaxy2025}.

\subsection{Subhalo abundance matching}
\label{sec:sham}
Subhalo abundance matching (SHAM) \cite{kravtsovDarkSideHalo2004, valeLinkingHaloMass2004, conroyModelingLuminositydependentGalaxy2006} is a widely used framework for establishing the galaxy–halo connection.
It is based on the assumption that there exists a monotonic relation between the mass of a (sub-)halo and the mass or luminosity of the galaxy it hosts.
In its simplest form, two catalogs covering the same sky area (one ranked by halo mass, one by luminosity or stellar mass) are matched in rank order, that is, the most massive halo is matched with the brightest (or most massive) galaxy, the second most massive halo with the second brightest galaxy, and so on.

SHAM was used in the past to assign realistic positions to galaxy catalogs sampled from a precursor of the GalSBI model \citepalias{bernerFastForwardModelling2024}.
The (sub-)halo catalog was generated the same way it is done in this work.
Additionally, a large galaxy catalog was sampled from the galaxy population model from \cite{herbelRedshiftDistributionCosmological2017}.

Since GalSBI consists of two separate galaxy populations (blue and red), also the halo catalog has to be split into two populations.
In \citetalias{bernerFastForwardModelling2024}, this was modelled via two quenching mechanisms:
mass quenching for central halos and environment quenching for subhalos.
Both were parametrized by step functions with characteristic mass limit and characteristic time scale since merger respectively.
The galaxy-halo connection can then be obtained as a function of redshift by matching the galaxy and corresponding halo catalog in small redshift bins.
For more details, we refer to \citetalias{bernerFastForwardModelling2024}.

\begin{figure}
    \centering
    \includegraphics[width=1\linewidth]{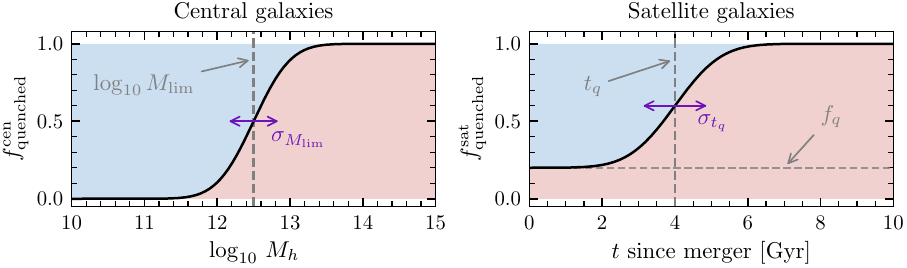}
    \caption{
    Illustration of the SHAM quenching model.
    Central galaxies are quenched based on halo mass, while satellite galaxies are quenched based on time since infall.
    Shaded regions show the red and blue galaxy fractions as a function of the respective quenching variable.
    Parameter values are chosen for illustration only.
    }
    \label{fig:quenching_illustration}
\end{figure}

Compared to \citetalias{bernerFastForwardModelling2024}, we change two aspects.
First, we relax the assumption of sharp quenching transitions, allowing for more flexible and realistic quenching behaviour parametrized by error functions $\mathrm{erf}$, while retaining the general approach of mass and environment quenching based on characteristic scales; see figure \ref{fig:quenching_illustration} for an illustration.
For central galaxies, the quenched fraction depends on halo mass,
\begin{equation} \label{eq:sham_centrals}
    f_\mathrm{quenched}^\mathrm{cen}(M_h) = \frac{1}{2}\left[1 + \mathrm{erf}\!\left(\frac{\log_{10} M_h - \log_{10} M_\mathrm{lim}}{\sqrt{2}\,\sigma_{M_\mathrm{lim}}}\right)\right],
\end{equation}
where $\mlim$ is the characteristic quenching mass and $\sigma_{M_\mathrm{lim}}$ controls the width of the transition.
For satellite galaxies, the red fraction depends on the time since the merger
\begin{equation} \label{eq:sham_sat}
    f_\mathrm{quenched}^\mathrm{sat}(t) = \fq + \frac{1 - \fq}{2}\left[1 + \mathrm{erf}\!\left(\frac{t - t_q}{\sqrt{2}\,\sigma_{t_q}}\right)\right],
\end{equation}
where $\tq$ is the characteristic quenching timescale, $\sigma_{t_q}$ the width of the transition, and $\fq$ the initial quenched fraction at infall.
The blue fractions follow as 
$f_\mathrm{blue} = 1 - f_\mathrm{quenched}$.
Based on these probabilities, all halos get tagged as hosting either a blue or a red galaxy.

Second, we use a more efficient SHAM scheme based on optimal transport \citep{fischbacherSHAMOTRapidSubhalo2025} to avoid repeated sampling and sorting of large galaxy catalogs.
Below, we give a short introduction into optimal transport and later its use for SHAM.

\subsection{Optimal transport}
\label{sec:ot}
We briefly review the theory of \emph{optimal transport (OT)} \citep[see][for review]{villaniOptimalTransport2009}, focusing on the discrete case.
Intuitively, OT provides a framework for finding the most efficient way to move mass from one distribution to another.

Consider two discrete multidimensional distributions $\mathbf{p} \in \mathbb{R}_+^m$ and $\mathbf{q} \in \mathbb{R}_+^n$, both normalized such that $\mathbf{p}^\mathrm{T} \mathbf{1}_m = \mathbf{q}^\mathrm{T} \mathbf{1}_n = 1$, where $\mathbf{1}$ is a vector of ones.
Define a \emph{cost matrix} $M \in \mathbb{R}_+^{m \times n}$, where $M_{ij}$ represents the cost of moving mass from $p_i$ to $q_j$.
The cost matrix $M$ must satisfy standard metric properties.
A \emph{transport plan} $\gamma \in \mathbb{R}_+^{m \times n}$ specifies how mass is moved from $\mathbf{p}$ to $\mathbf{q}$, satisfying
\begin{equation}
\gamma \mathbf{1} = \mathbf{p}, \quad \gamma^\mathrm{T} \mathbf{1} = \mathbf{q}, \quad \gamma \ge 0,
\end{equation}
that is, $\gamma_{ij}$ is the amount of mass transported from $p_i$ to $q_j$.

The \emph{optimal transport solution} $\gamma^*$ is the transport plan that minimizes the total cost:
\begin{equation}
\gamma^* = \arg\min_{\gamma \in \mathbb{R}_+^{m \times n}} \sum_{i,j} \gamma_{ij} M_{ij}.
\end{equation}
This problem can be formulated as a linear program and solved using standard solvers. The associated minimal cost defines the \emph{optimal transport distance} (or \emph{Wasserstein distance}) between $\mathbf{p}$ and $\mathbf{q}$.

In general, the solution may not be unique. To obtain smoother solutions, one can use \emph{regularized optimal transport}, most commonly \emph{entropic regularization} \cite{cuturi2013sinkhorn}:
\begin{equation} \label{eq:reg_ot}
\gamma^* = \arg\min_{\gamma \in \mathbb{R}_+^{m \times n}} \sum_{i,j} \gamma_{ij} M_{ij} + \varepsilon \sum_{i,j} \gamma_{ij} \log \gamma_{ij},
\end{equation}
where $\varepsilon > 0$ controls the strength of the regularization.
This ensures a unique, smooth solution that can be efficiently computed using the \emph{Sinkhorn algorithm}.

\subsection{SHAM-OT}
\label{sec:sham-ot}
\citetalias{fischbacherSHAMOTRapidSubhalo2025} demonstrated that OT can be used to reformulate SHAM in an efficient way.
Instead of creating large halo and galaxy catalogs, we can use normalized histograms of halo masses and galaxy luminosities, achieving a computational speedup of $O(100)$.
These histograms act as the two discrete distributions $\mathbf{p}$ and $\mathbf{q}$, and the OT solution provides a mapping from halo mass to absolute magnitude (or stellar mass).
In the limit of infinitely small histogram bins, the OT solution is mathematically equivalent to the traditional SHAM solution.

Scatter in the galaxy–halo connection is incorporated through entropic regularization, which smooths the transport plan.
The shape of this scatter depends on the choice of cost matrix.
By designing the cost matrix appropriately, specific scatter behaviors such as lognormal scatter in luminosity (similar to the lognormal scatter in stellar mass presented in \cite{behrooziAverageStarFormation2013}) can be emulated.
Unlike traditional SHAM, this does not require deconvolution techniques to preserve the luminosity function, since this is a fundamental property of the OT solution by construction.

Within our framework, we apply this methodology starting from a full-sky (sub-)halo catalog (see section \ref{sec:pinocchio}).
From this catalog, we precompute two products: a 2-dimensional histogram in redshift and halo mass, and a suite of small per-image catalogs containing only the sources within each image boundary. 
For each redshift bin, the 1-dimensional halo mass histogram is matched with a galaxy histogram using OT.
The galaxy histogram is computed on the fly from the analytical luminosity function at the corresponding redshift, scaled to full-sky.
This yields a mapping from halo mass $M_\mathrm{h}$ to absolute magnitude $m$ at every redshift, which is applied to the per-image catalogs.
These catalogs already contain positions and redshifts for each source; the mapping adds the corresponding absolute magnitude.
SEDs and morphologies are then assigned following the procedure of previous GalSBI modelling. 
Details on the position assignment procedure and a validation that the \pinocchio mass limit does not bias our inference are given in appendix~\ref{app:posassignment}.

\subsection{Clustering statistics}
\label{sec:cl}
We compare the clustering signal in our simulation with the data using the angular power spectrum $C_\ell$.
As shown in appendix \ref{app:cl_norm}, the mask, varying survey depth, and shot noise are consistently forward modelled in our simulations, so power spectra from simulations and data covering the same footprint can be compared directly, without separately correcting for these effects.
However, this is no longer true when comparing power spectra estimated from different footprints, since the mask, depth, and shot noise differ between them.
During inference, we compare simulated power spectra estimated from small patches to the data power spectrum estimated from the full footprint (see section \ref{sec:sbi} below).
This comparison between different footprints is only possible if we correctly mitigate the effects of mask, depth, and shot noise.

The contribution of varying survey depth to the power spectrum is characterized using simulations of the full DES~Y3 footprint with \ufig and the GalSBI model from \citetalias{fischbacherGalSBIPhenomenologicalGalaxy2025} with positions being sampled from a uniform distribution.
The raw power spectra of these simulations nicely capture spikes in the power spectra of the data due to varying depth of the DES survey due to survey strategy and CCD patterns; see appendix \ref{app:cl_norm}.

For the simulations including clustering, we estimate $C_\ell$ of our simulations using the pseudo-$C_\ell$ method implemented in \texttt{NaMaster} \citep{alonsoUnifiedPseudo$C_ell$Framework2019}.
We project the galaxy catalog after our selection onto a \texttt{HEALPix} map \citep{gorskiHEALPixFrameworkHigh2005} with $NSIDE=1024$.
The overdensity field is then computed as
\begin{equation}
    \delta(\hat{\mathbf{n}}) = \frac{n_\mathrm{g}(\hat{\mathbf{n}})}{\alpha \, \bar{n}_\mathrm{r}(\hat{\mathbf{n}})} - 1,
\end{equation}
where $n_\mathrm{g}(\hat{\mathbf{n}})$ is the galaxy count in pixel $\hat{\mathbf{n}}$, 
$\bar{n}_\mathrm{r}(\hat{\mathbf{n}})$ is the mean galaxy count from random realizations generated as described above
and $\alpha = N_\mathrm{g} / N_\mathrm{r}$ is the normalization factor with $N_\mathrm{g}$ and 
$N_\mathrm{r}$ the galaxy counts from the simulation and the random realizations, respectively.
A mean subtraction is applied to remove the monopole contribution.

To mitigate edge effects from the survey geometry, we apply a $C^2$ apodization mask with a smoothing scale of $\SI{1}{\degree}$ to the survey footprint.
The pseudo-power spectrum $\tilde{C}_\ell$ is computed from the masked field and corrected for mode-coupling effects via the coupling matrix $M_{\ell\ell'}$:
\begin{equation}
    C_\ell = \sum_{\ell'} M^{-1}_{\ell\ell'} \tilde{C}_{\ell'}.
\end{equation}
The estimated power spectrum is linearly binned in multipole space with bin width $\Delta\ell$. 
We subtract the shot noise contribution
\begin{equation}
    N_\ell = \frac{\Omega_\mathrm{survey}}{N_\mathrm{g}},
\end{equation}
where $\Omega_\mathrm{survey}$ is the effective survey area in steradians.

\subsection{Simulation-based inference}
\label{sec:sbi}
We adopt the same inference strategy as in \citetalias{fischbacherGalSBIPhenomenologicalGalaxy2025} based on an iterative implementation of Approximate Bayesian Computation (ABC).
We start by sampling points from the prior distribution, simulate a patch of the sky, and compute distances between simulations and data.
We then accept the fraction of points with the lowest distance to approximate our posterior.
This posterior is then used in the next iteration as prior and we simulate a new previously unseen patch of the sky.
In the following, we will give an overview of our prior distribution, the newly adapted emulator, the choice of distance measures and the tile selection.

\subsubsection{Prior}
Compared to the GalSBI model in \citetalias{fischbacherGalSBIPhenomenologicalGalaxy2025}, we add seven additional parameters to the model related to the galaxy-halo connection.
Five of them are connected to the split into quenched and unquenched hosts: $\mlim, \sigmlim, \tq, \sigtq$ and $\fq$; see Equations \ref{eq:sham_centrals} and \ref{eq:sham_sat}.
Additionally, we use entropic OT for the SHAM solution, where the strength of the scatter is controlled by regularization parameters (one per population): $\varepsilon_\mathrm{blue}$ and $\varepsilon_\mathrm{red}$.
For the cost of optimal transport, we use the proposed solution from \citetalias{fischbacherSHAMOTRapidSubhalo2025}.

We constrain all 60~parameters simultaneously.
However, since we have strong prior knowledge on the parameters of the galaxy population parameter from previous work, we use the posterior of \citetalias{fischbacherGalSBIPhenomenologicalGalaxy2025} as prior for the galaxy population parameters.
We have tested the impact of the prior and find consistent results if the prior on the galaxy population parameters is increased.
A summary of the prior is given in Table \ref{tab:prior}.

\begin{table}[]
    \centering
    \begin{tabular}{lllll}
        \toprule
        Parameter & Definition & Prior\\
        \midrule
        Galaxy population model & Section 2.1 of \citetalias{fischbacherGalSBIPhenomenologicalGalaxy2025} & posterior of \citetalias{fischbacherGalSBIPhenomenologicalGalaxy2025} \\
        $\log_{10} \mlim$ & Equation \ref{eq:sham_centrals} & $\mathcal{U}[11.5,13.5]$ \\
        $\sigmlim$ & Equation \ref{eq:sham_centrals} & $\mathcal{U}[0.001,1]$ \\
        $\tq$ & Equation \ref{eq:sham_sat} & $\mathcal{U}[0,4]$ \\
        $\sigtq$ & Equation \ref{eq:sham_sat} & $\mathcal{U}[0.001,0.5]$ \\
        $\fq$ & Equation \ref{eq:sham_sat} & $\mathcal{U}[0,0.5]$ \\
        $\log_{10}\varepsilon_\mathrm{blue}$ & Equation \ref{eq:reg_ot} & $\mathcal{U}[-3.5,-1]$ \\
        $\log_{10}\varepsilon_\mathrm{red}$ & Equation \ref{eq:reg_ot} & $\mathcal{U}[-3.75,-1]$ \\
        \bottomrule
   \end{tabular}
    \caption{
    We indicate for each set of parameters where they are defined in the text and give the prior.
    $\mathcal{U}$ denotes uniform distributions.
    }
    \label{tab:prior}
\end{table}

\subsubsection{Emulator}
\label{sec:emulator}
In \citetalias{fischbacherGalSBIPhenomenologicalGalaxy2025}, a two-stage emulator was used, modelling the detection probability and the photometric noise separately.
While blending was already one of the main challenges there, its impact on the detection classifier could be mitigated because blending was a spatially uniform nuisance: with random galaxy positions, the blending risk depends only on the galaxy population, not on location within the image.
It was therefore sufficient to preserve the 1-point function of galaxies.
If a galaxy was missed due to blending, an equivalent detection elsewhere left the galaxy distribution unchanged.

With clustered positions, however, this mitigation strategy fails. 
Blending becomes locally more severe in overdense regions, making the detection probability a function of local environment rather than a global image property.
Furthermore, a missed detection in a dense region cannot be compensated by an additional detection in a sparse one without biasing the clustering statistics.
A detection classifier without any local blending information might therefore bias the clustering statistics.
We confirmed this by testing the old emulator setup, finding significant deviations in the angular power spectrum.

Although local environment features can be passed to the classifier to partly mitigate this \citep{zhangEmulatingRedshiftMixing2026}, we avoid this problem entirely by using image simulations for the detection step.
When using \sextractor in forced photometry mode, the detection probability is purely determined by the band used for detection, in our case the $i$-band.
We therefore only simulate the $i$-band and run \sextractor on it.
Photometry in the remaining bands is then emulated using the same normalizing flow architecture as in \citetalias{fischbacherGalSBIPhenomenologicalGalaxy2025}.
This still yields a considerable speed-up by avoiding 4 out of 5 image simulations and source extractions, while remaining unbiased with respect to blending.
We validate this approach by comparing clustering statistics from an emulator-generated full DES~Y3 simulation against those from full image simulations across different magnitude and color cuts.
In contrast to the old emulator setup, we now find almost perfect agreement.

\subsubsection{Tile selection}
\begin{figure}
    \centering
    \includegraphics[width=1\linewidth]{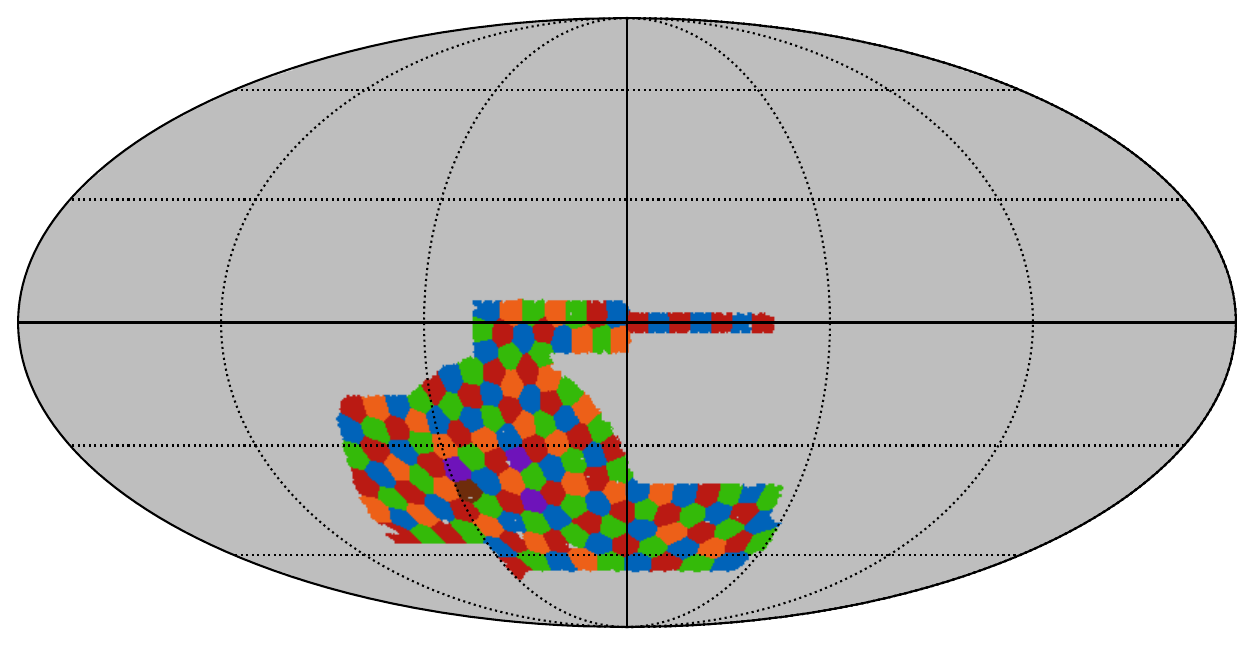}
    \caption{
    Footprint of the DES~Y3 survey and the separation in patches used in this work.
    The full survey footprint is split into 150~patches of roughly equal size.
    During inference, one patch is used per iteration.
    }
    \label{fig:des_footprint}
\end{figure}

At each iteration of the inference, we measure the angular power spectrum for which we need a large enough survey area to obtain stable $C_\ell$ measurements.
At the same time, the computational cost of one iteration should be as minimal as possible.
We achieve this by dividing the survey footprint into 150~spatially compact patches of about $\SI{30}{\square\deg}$.
The split is performed using K-means clustering on the 3-dimensional Cartesian coordinates of the unmasked \texttt{HEALPix} pixels; see figure \ref{fig:des_footprint} for the different patches.

\subsubsection{Distance measures}
Compared to \citetalias{fischbacherGalSBIPhenomenologicalGalaxy2025}, we adapt the distance measures as follows.
Since we use DES Y3 data instead of HSC, no reliable redshift information is available.
We therefore drop all redshift-based summary statistics.
We thus keep the fractional difference in number of galaxies $d_\mathrm{frac}$ and a weighted Wasserstein distance $W_{1,\mathrm{w}}$ for magnitudes and flux fractions in all bands as well as flux radius and absolute ellipticities in the $i$-band.
For details on the computation of $d_\mathrm{frac}$ and $W_{1,\mathrm{w}}$, we refer to \citetalias{fischbacherGalSBIPhenomenologicalGalaxy2025}.

To include an informative distance measure for the clustering statistics, we add a distance based on the angular power spectrum.
The angular power spectrum is computed on 12~galaxy sub-samples defined by magnitude and color selections.
In magnitude, we split the galaxy sample using $i$-band \texttt{MAG\_AUTO} into four bins: $[17,23.5]$, $[21, 22]$, $[22, 23]$, and $[23, 23.5]$.
In color, we divide the sample into a \enquote{bluer} and a \enquote{redder} sub-sample using a linear boundary in the $(g-r,\ r-i)$ color--color plane.
Galaxies satisfying
\begin{equation}
    \label{eq:redder_bluer}
    -1.12\,(g-r) - 1.18\,(r-i) + 2.3 \geq 0
\end{equation}
are assigned to the bluer sample.
The coefficients are derived from a linear SVC fit to the true red and blue populations in GalSBI.
We further found that this boundary provides a good separation between star-forming and quiescent galaxies in COSMOS2020 \citep{weaverCOSMOS2020PanchromaticView2022} data based on specific star formation rates from LePhare \citep{arnoutsMeasuringModellingRedshift1999,ilbertAccuratePhotometricRedshifts2006} and EAZY \cite{brammerEAZYFastPublic2008}.
Note that the precise form of this boundary does not critically affect the inference, since the same split is applied consistently to both simulations and data; nevertheless, a well-chosen boundary will improve convergence.
Combining the three color selections (\enquote{all}, \enquote{bluer}, \enquote{redder}) with the four magnitude bins yields 12 sub-samples in total.

For each sub-sample, we compute the angular power spectrum $C_\ell^\mathrm{sim}$ as described in section \ref{sec:cl} with $\ell$-range from $[0, 2000]$ and bin width $\Delta \ell=100$.
The distance is then
\begin{equation}
d_{C_\ell} = \Delta C_\ell^T \, \Sigma^{-1} \, \Delta C_\ell
\quad \text{with} \quad
\Delta C_\ell = \log_{10}C_\ell^\mathrm{sim} - \log_{10}C_\ell^\mathrm{data},
\end{equation}
where $C_\ell^\mathrm{data}$ is the mean angular power spectrum in the data across all patches and $\Sigma$ is estimated as the variance of $C_\ell$ across the 150~patches in the data.

Finally, the individual distances are combined as in \citetalias{fischbacherGalSBIPhenomenologicalGalaxy2025} with the following weights:
\begin{equation}
    \label{eq:final_distance}
    \delta = 0.2 d_\mathrm{frac} + 0.4 W_{1,\mathrm{w}} + 0.4 d_{C_\ell}.
\end{equation}
The weights are chosen similarly to previous work \citep{moserSimulationbasedInferenceDeep2024,fischbacherGalSBIPhenomenologicalGalaxy2025} but halving the photometric distance measure to incorporate the clustering-based distance measure.

\section{Results}
\label{sec:results}
After completing inference, we obtain a posterior distribution of the 60-dimensional model parameter space of the galaxy population model.
To generate image simulations, we draw a single point from this posterior and use it to sample the galaxy catalog.

\subsection{Galaxy population and galaxy-halo connection constraints}
\label{sec:galpop_res}
The constrained galaxy population model can not only be used to generate realistic galaxy catalogs, but also represents a forward-modelling-based measurement of quantities like the luminosity function or the galaxy–halo connection.
The posterior distributions of the luminosity function parameters are highly consistent with those obtained in \citetalias{fischbacherGalSBIPhenomenologicalGalaxy2025}.
Since the resulting B-band luminosity function is therefore essentially unchanged compared to \citetalias{fischbacherGalSBIPhenomenologicalGalaxy2025}, our measurement remains consistent with previous forward-modelling-based measurements \citep{tortorelliMeasurementBbandGalaxy2020,moserSimulationbasedInferenceDeep2024} and, through the comparisons performed in those works, with the range of external measurements \citep{giallongoBBandLuminosityFunction2005,ilbertVIMOSVLTDeepSurvey2006,zuccaZCOSMOSSurveyRole2009,lovedayGalaxyMassAssembly2012,coolGalaxyOpticalLuminosity2012,beare$z12$Optical2015}.

For the quenching parameters, we find consistent results compared to
\citetalias{bernerFastForwardModelling2024}; see appendix \ref{sec:galpop_gh} for the posterior contours.
They found the mass limit $\mlim$ and the characteristic quenching time scale $\tq$ to be mutually degenerate, with fiducial values of $\mlim=\SI{8e12}{\per\hubble \solarmass}$ and $\tq=\SI{2}{\giga\year}$.
Our measurements are less degenerate and are consistent with the posterior distribution reported there, but with somewhat lower $\mlim$ and higher $\tq$.
We further note that the values are not directly comparable, as we relaxed the assumption of a sharp threshold.
The newly introduced quenching parameters are all mildly constrained and do not approach any prior boundaries.

\begin{figure}
    \centering
    \includegraphics[width=1\linewidth]{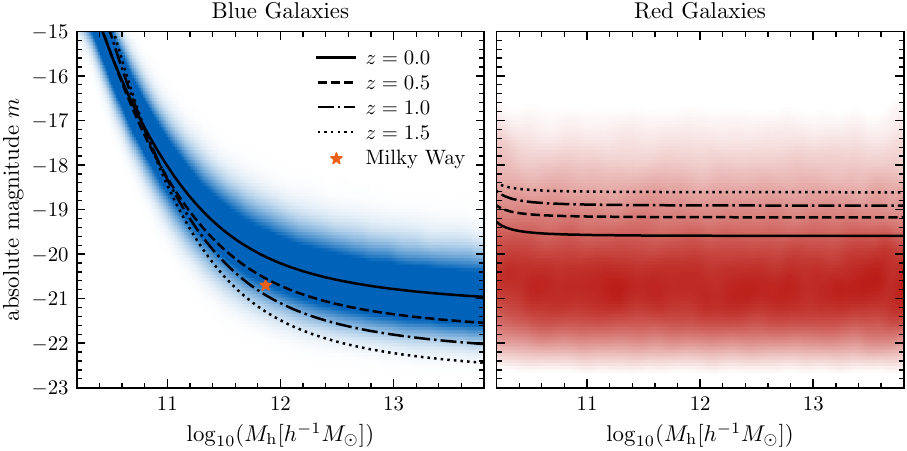}
    \caption{
    Measured galaxy-halo connection for red and blue populations, their scatter and redshift evolution.
    The shaded region shows the galaxy-halo connection at $z=0$ including the scatter.
    The different lines indicate the mean absolute magnitude at given halo mass at different redshifts.
    Note that due to the large scatter for red galaxies, the magnitude distribution for a given halo mass is not symmetric and therefore the mean magnitude does not correspond to the peak of the distribution.
    We further show a measurement of the Milky Way \cite{bland-hawthornGalaxyContextStructural2016} for comparison.
    }
    \label{fig:transport_plan}
\end{figure}

The scatter amplitude parameters, which set the strength of the regularization in the OT solution, are both tightly constrained.
Figure~\ref{fig:transport_plan} shows the inferred galaxy-halo connection, including scatter, for a randomly chosen draw of the posterior.
The shaded region shows the measured scatter of the galaxy-halo connection, not the posterior uncertainty on the connection itself.
Our results align well with measurements of close-by blue galaxies such as the Milky Way and Andromeda \citep[e.g.,][]{berghLocalGroupGalaxies1999,tempelDustcorrectedSurfacePhotometry2010,penarrubiaDynamicalModelLocal2014,bland-hawthornGalaxyContextStructural2016,watkinsEvidenceIntermediateMassMilky2019}, local elliptical galaxies \cite{faberMassesMasstolightRatios1979}, and measurements using the Tully-Fisher relation \cite{rijckeGeneralizationsTullyFisherRelation2007} and the fundamental plane \cite{welMasstoLightRatiosField2005}.
The measured scatter is significantly larger for red galaxies than for blue ones, which we mainly attribute to the fact that satellite galaxies are known to exhibit larger scatter than central galaxies \citep[e.g.][]{englerDistinctStellartohaloMass2020} and that satellites dominate the red population, whereas centrals dominate the blue one.

\subsection{Validation}
\label{sec:visual}
\begin{figure}
    \centering
    \includegraphics[width=1\linewidth]{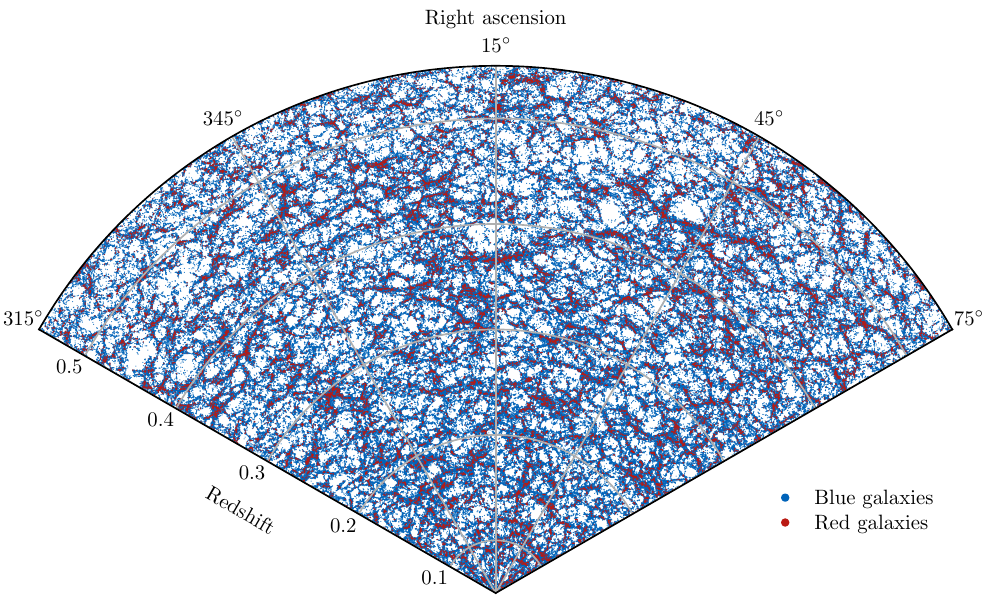}
    \caption{
    Red and blue galaxies in a DES simulation at declination $\SI{-51.25}{\degree}<\delta<\SI{-51.75}{\degree}$.
    For illustrative reasons we only show galaxies up to $z<0.55$ and a $\SI{60}{\degree}$ wedge in right ascension.
    }
    \label{fig:wedge}
\end{figure}

\begin{figure}
    \centering
    \includegraphics[width=1\linewidth]{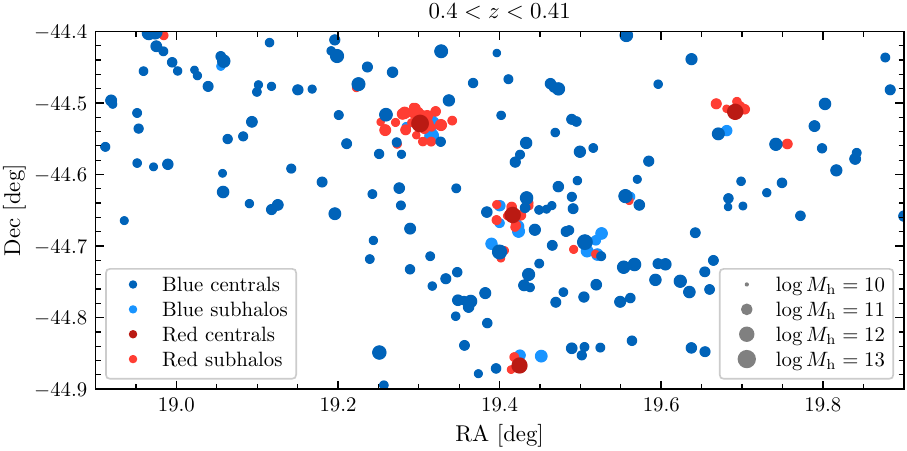}
    \caption{
    Galaxy distribution in right ascension and declination within a redshift slice of $0.4<z<0.41$.
    The population consists of four subpopulations: Blue centrals, blue subhalos, red centrals, red subhalos.
    The mass of the halos is illustrated as the size of the dots.
    }
    \label{fig:radec_slice}
\end{figure}

In this part, we validate the model using DES and HSC imaging data (see Figure~\ref{fig:images} for an example of the corresponding image simulations).
To illustrate the clustering of red and blue galaxies, figure~\ref{fig:wedge} shows a declination slice through a DES simulation.
Red galaxies preferentially occupy high-density regions, while blue galaxies are more spatially dispersed.
In figure~\ref{fig:radec_slice}, we show a small redshift slice of galaxies split into blue and red as well as centrals and subhalos.
As expected from our quenching model, massive red centrals reside in the most massive halos whereas blue centrals occupy the smaller mass halos.

In the following subsections, we validate the simulations more quantitatively, comparing photometric catalog properties, angular power spectra $C_\ell$, and redshift distributions $n(z)$ for DES and HSC separately.

\subsubsection{DES}
\label{sec:results_des}
\begin{figure}
    \centering
    \includegraphics[width=1\linewidth]{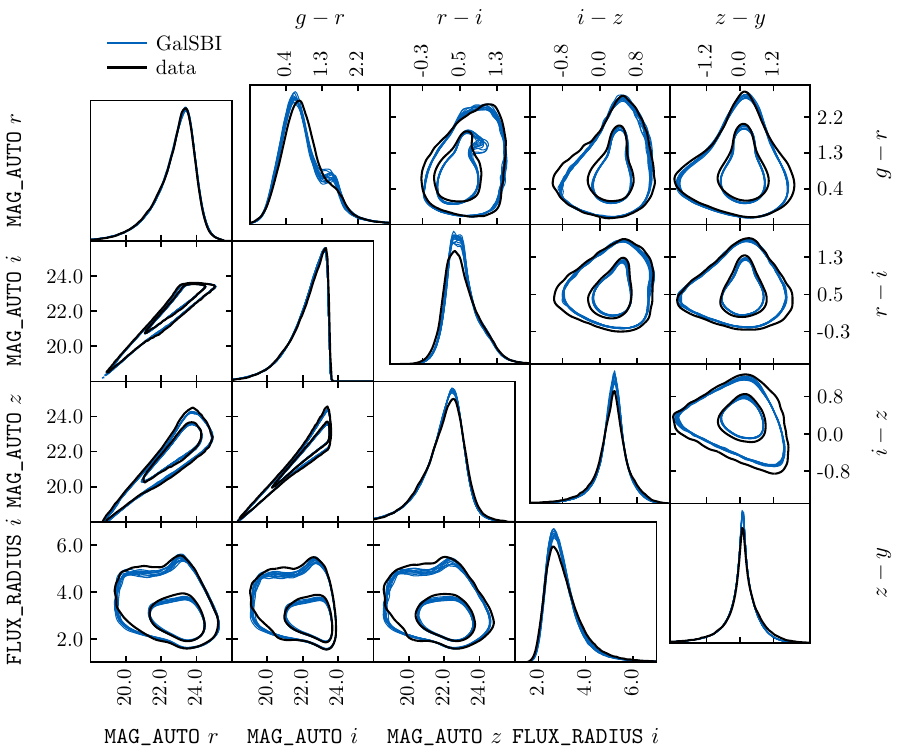}
    \caption{
    Comparison of magnitudes, sizes, and colors between 12~DES Y3 simulations and data.
    }
    \label{fig:des_photo}
\end{figure}

We simulate 12~full DES Y3 footprints using 12~independent posterior draws from the galaxy population model, cycling through the 4~rotational realizations of the halo catalog.
The ensemble therefore captures both galaxy population model uncertainty and, partially, sample variance from the halo catalog.
In figure \ref{fig:des_photo}, we show the magnitude, color and size distributions of our simulations compared to data.
Our DES simulations agree very well with the data, also in parameters not shown in the figure such as shape or surface brightness, and at a level comparable to simulations using the model from \citetalias{fischbacherGalSBIPhenomenologicalGalaxy2025}.
This confirms that introducing clustering does not degrade the modelling of photometry and morphology.

\begin{figure}
    \centering
    \includegraphics[width=1\linewidth]{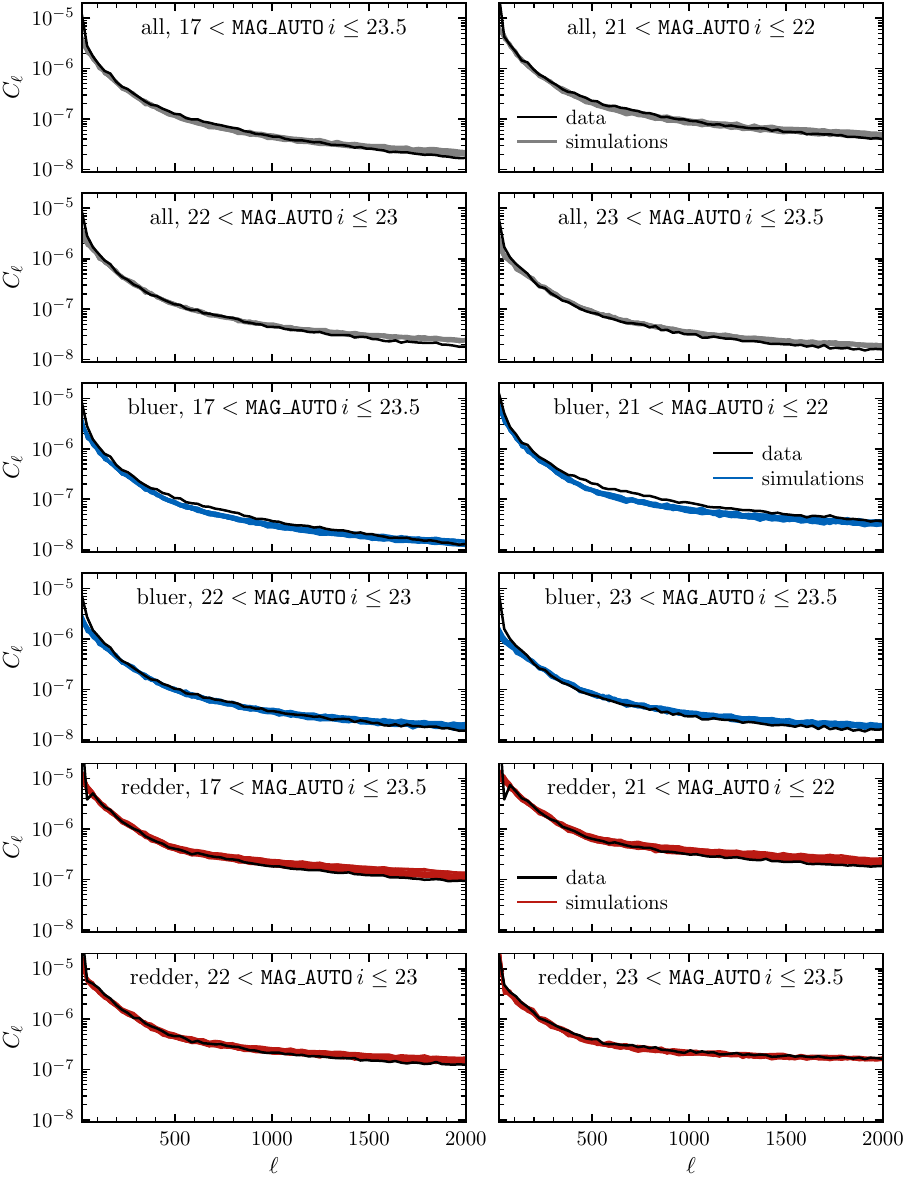}
    \caption{
    Comparison of angular power spectra of galaxy positions between data and simulations across different cuts in magnitude and galaxy samples.
    The bluer and redder sample are split using the photometric color selection defined in equation \ref{eq:redder_bluer}.
    }
    \label{fig:cl_comparison}
\end{figure}

Since galaxy positions are now clustered, we can further compare angular power spectrum of galaxy positions, which is shown in figure \ref{fig:cl_comparison}.
We find good agreement across different magnitude and color cuts.
The data power spectra lie within the 12~simulations, except for the bright bluer samples, where our simulations lack a bit of power.

\subsubsection{HSC~DUD}
For HSC~DUD, we create 30~simulations of the region that overlaps with the COSMOS field.
Each simulation uses a different draw from the galaxy population model and a different rotational realization of the halo simulation; see appendix \ref{app:rotation} for details.
Figure \ref{fig:hsc_photo} shows the analogous comparison from figure \ref{fig:des_photo} in DES, now on HSC~DUD data.
While the depth is very different to the DES data, the agreement between data and simulation is very similar.
This demonstrates that GalSBI can accurately reproduce galaxy populations across different depth and image systematics regimes.
We note that compared to \citetalias{fischbacherGalSBIPhenomenologicalGalaxy2025}, we see larger variability between the different simulations, mainly visible in the colors (for example at $g-r\sim1.5$).
This is due to the effect of sample variance, which was not included in \citetalias{fischbacherGalSBIPhenomenologicalGalaxy2025}.
The larger scatter compared to the DES~simulations (figure \ref{fig:des_photo}) can be attributed to the much smaller validation area, which is roughly $\SI{2}{\squaredegree}$ compared to almost $\SI{5000}{\squaredegree}$ for DES.

\begin{figure}
    \centering
    \includegraphics[width=1\linewidth]{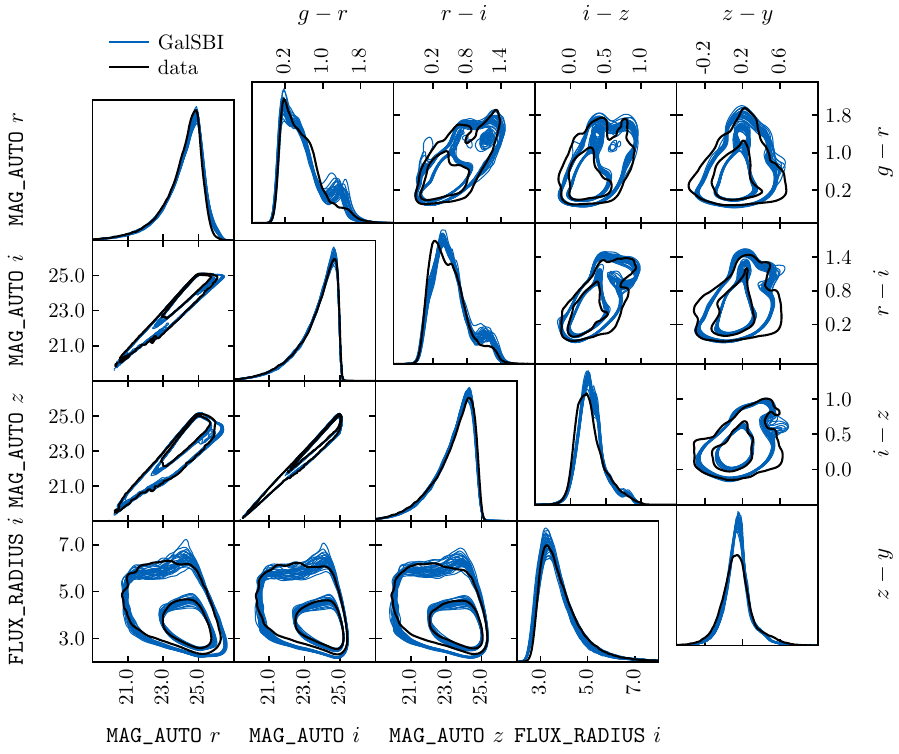}
    \caption{
    Comparison of magnitudes, sizes and colors between 30~HSC DUD simulations and the data.
    }
    \label{fig:hsc_photo}
\end{figure}

\begin{figure}
    \centering
    \includegraphics[width=1\linewidth]{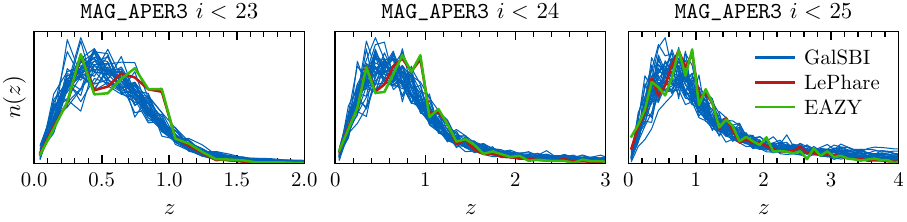}
    \caption{
    Comparison of redshift distributions for different magnitude cuts between simulation and data.
    The photometric redshifts for the data are estimated using LePhare and EAZY.
    }
    \label{fig:hsc_redshift}
\end{figure}

We can further compare our predicted $n(z)$ distributions with photometric redshifts estimated from LePhare and EAZY, see figure \ref{fig:hsc_redshift}.
While in previous work, our redshift distributions were much smoother than LePhare and EAZY, with the newly introduced clustering, we obtain similarly spiky redshift distributions as those of the real data.

\label{sec:results_hsc}

The mean redshifts across three magnitude cuts agree well with LePhare (LP) and EAZY (E), given as 
\begin{equation}
    \begin{aligned}
        &\bar{z}_\mathrm{GalSBI,<23} = 0.575 \pm 0.014,   
        &\bar{z}_\mathrm{LP,<23} = 0.604 \pm 0.012,   
        &&\bar{z}_\mathrm{E,<23} = 0.617 \pm 0.012,\\
        &\bar{z}_\mathrm{GalSBI,<24} = 0.788 \pm 0.023,  
        &\bar{z}_\mathrm{LP,<24} = 0.823 \pm 0.019,   
        &&\bar{z}_\mathrm{E,<24} = 0.841 \pm 0.019,\\ 
        &\bar{z}_\mathrm{GalSBI,<25} = 1.100 \pm 0.040,   
        &\bar{z}_\mathrm{LP,<25} = 1.109 \pm 0.019,   
        &&\bar{z}_\mathrm{E,<25} = 1.108 \pm 0.018,  
    \end{aligned}
\end{equation}
where mean and uncertainty from LePhare and EAZY are estimated using the reweighted mocks from \cite{moserSimulationbasedInferenceDeep2024}.

We can further estimate which part of the uncertainty stems from sample variance in the relatively small COSMOS field and which part is uncertainty in the GalSBI model propagated into the redshift distribution.
For this, we repeat the 30~simulations twice, once with fixed draw from the galaxy population model and varying halo catalogs, once with fixed halo catalog and varying galaxy population model.
From this, we can estimate the redshift uncertainties as 
\begin{equation}
    \begin{aligned} 
        \bar{z}_\mathrm{GalSBI,23} &= 0.575 
            \pm 0.005\,\text{(model)} 
            \pm 0.012\,\text{(sample)}\\
        \bar{z}_\mathrm{GalSBI,24} &= 0.788 
            \pm 0.010\,\text{(model)} 
            \pm 0.016\,\text{(sample)}\\
        \bar{z}_\mathrm{GalSBI,25} &= 1.100 
            \pm 0.026\,\text{(model)} 
            \pm 0.022\,\text{(sample)}\\
    \end{aligned}
\end{equation}
Sample variance is more dominant in the brighter cuts, which probe a smaller volume at lower redshift.
We find that for the size of the COSMOS field, sample variance is the dominant factor in the two brighter cuts $<23$ and $<24$ and subdominant, yet still significant contributor in the faintest sample.
More details on the computation of these uncertainties and plots of the redshift distributions of the other two simulations suites are given in appendix \ref{app:nz}.

The agreement of the redshift distribution varies across magnitude bins.
In the faint sample, the agreement is very good, within $0.2\sigma$ assuming uncertainties add in quadrature.
At the brighter end, GalSBI redshifts are slightly lower than both LePhare and EAZY, with discrepancies of $1.2$ to $1.6\sigma$
relative to LePhare and slightly larger relative to EAZY.
We note that there is also a disagreement between LePhare and EAZY themselves ($0.8\sigma$ for the brightest cut).
The discrepancies are therefore largest in the sample-variance-dominated cuts, where the photo-$z$ codes themselves agree least, and negligible in the population-model-dominated cut, where they mutually agree.
Whether this mild discrepancy can be attributed to a sample-variance effect in the photo-$z$ codes or to imperfect galaxy population modelling at the bright end will be investigated in future work.
\section{Conclusion}
\label{sec:conclusion}

We have presented an update to the GalSBI model, which jointly constrains the galaxy population and the galaxy-halo connection.
Using the constrained model to generate realistic DES and HSC DUD data, we accurately reproduce a wide range of observables including luminosity functions, photometry, morphology, redshift distributions and -- for the first time within the GalSBI framework -- the galaxy-halo connection and the angular power spectrum of galaxy positions across magnitude and color selections.
The updated GalSBI model including the SHAM-OT-based position assignment is publicly available in the \galsbi Python package \citep{fischbacherGalsbiPythonPackage2025}.

During inference, we used an emulator setup to reduce the computational cost of our analysis.
Unlike previous work with uniformly distributed positions, we found that treating blending as an averaged effect biases the clustering statistics.
We therefore replaced the detection classifier with image-simulation-based detection, combined with a normalizing flow to predict photometric and morphological measurements in the remaining bands. 
This approach achieves high accuracy in modelling the detection probability while still providing a significant speedup by avoiding image simulations and source extraction in all bands.

This new model can now be applied to a variety of tasks.
Shape and blending calibration will be possible with higher accuracy since the image simulations now contain clustered positions.
For blending, we find that the clustering results in almost no reduction in the number of detected galaxies for DES, but $\sim 7\%$ reduction for the HSC deep fields.
This suggests that including clustering is important for calibrating deep Stage-IV data.

In a cosmic shear setup, we can now consistently account for source clustering in our simulations.
Furthermore, our redshift distributions are subject to sample variance, and we can easily quantify its impact.
For the size of the COSMOS field, we showed that sample variance is the dominant source of uncertainty on the mean redshift over galaxy population uncertainty for faint galaxies with $i$-band magnitude below 24.
Detailed investigation of how calibration on the small COSMOS field may induce biases in the final analyses of photometric redshifts is left for future work.

Clustered GalSBI simulations could be used to populate large simulation suites with varying cosmology such as the CosmoGrid \citep{kacprzakCosmoGridV1Simulated$w$CDM2023} enabling cosmic shear, galaxy clustering and their correlation analyses, including beyond two-point statistics \citep[e.g.][]{thomsenDarkEnergySurvey2026}.
However, this requires further validation of different components of the SHAM-pipeline.
The robustness of our halo and subhalo catalog generation, based on \pinocchio and the subhalo extraction of \citetalias{bernerFastForwardModelling2024}, can be tested against high resolution N-body simulations \citep{springelGADGETCodeCollisionless2001,springelCosmologicalSimulationCode2005,Potter:2016ttn,springelSimulatingCosmicStructure2021,garrisonAbacusCosmosSuite2018} and their speed and accuracy can be further benchmarked against particle mesh codes \citep[e.g., Disco-DJ][]{listDISCODJIIDifferentiable2025}.
The galaxy-halo connection modelling can be compared to existing approaches, for example, in the Euclid flagship simulation \citep{collaborationEuclidFlagshipGalaxy2024} that uses a combination of SHAM and halo occupation distribution to build the galaxy catalog \citep{carreteroAlgorithmBuildMock2015}.
Furthermore, our current scatter model (parametrized by the OT cost matrix) was developed to create symmetric and constant scatter for relatively low scatter amplitude.
This might need to be revisited given the large measured scatter for red galaxies.
An adapted cost matrix could also emulate varying scatter strength, for example, to account for higher scatter for subhalos at low mass and lower scatter for central galaxies at higher mass.
To apply GalSBI simulations to map-level or field-level analysis, the clustering statistics should be further validated on non-Gaussian statistics \citep[see e.g.][and references therein]{collaborationParameterMaskedMockData2024}.

\ufig image simulations have been validated on DES and HSC data, which were used to constrain the galaxy population model; a logical next step is to test and potentially refine the model on a survey with broader wavelength coverage, for example, the KiDS-VIKING dataset \citep{wrightFifthDataRelease2024,edgeVISTAKilodegreeInfrared2013}.
While preliminary results on this dataset suggest that extrapolating the current template parameters to the VISTA filters still yields accurate color distributions, constraining the model using the 9-band photometry of KiDS-VIKING will help tighten the model constraints.
However, the infrared VIKING filters introduce new image systematics that require careful validation and potential adaptation of the \ufig pipeline.

Such a refined and thoroughly validated GalSBI model will be ready to be used in Stage-IV cosmology.
Applying GalSBI to Rubin-LSST can build on existing HSC image simulation workflows, while generating Euclid image simulations will require additional validation to accurately model space-based imaging characteristics, particularly the PSF.

The extensions described above are not limited to GalSBI but apply equally to GalSBI-SPS \citep{tortorelliGALSBISPSStellarPopulation2025b}.
The SHAM workflow presented here can be easily adapted to GalSBI-SPS by matching the halo mass functions with the stellar mass functions instead of the luminosity function.
With GalSBI-SPS, this approach could be further extended by using the merger history to inform the star formation history (SFH) of galaxies.
This is of particular interest given that fast models of the statistical connection between SFH and the mass assembly history of dark matter halos have recently emerged in the literature \citep{alarconDiffstarPopGenerativePhysical2025}.

Beyond imaging surveys, GalSBI can be extended to spectroscopic data.
Both GalSBI and GalSBI-SPS can generate galaxy spectra, with GalSBI-SPS in particular having the potential to generate highly realistic ones.
Combined with the spectra simulator \texttt{USpec} \citep{fagioliForwardModelingSpectroscopic2018,fagioliSpectroimagingForwardModel2020,lucatortorelliUSpec2UltrafastSpectrainprep.}, spectroscopic surveys such as DESI \citep{desicollaborationDESIExperimentPart2016}, 4MOST \citep{dejong4MOSTProjectOverview2019} or PFS \cite{takadaExtragalacticScienceCosmology2014} can be used to constrain the model as well as to forward model the effect of different selection functions in these surveys.
In addition to fast spectra simulations, this would require SPS-based spectral emulation \citep{alsingSPECULATOREmulatingStellar2020,melchiorAutoencodingGalaxySpectra2023}, complementing existing magnitude emulation \citep{tortorelliProMageFastGalaxy2025}.
Combining both photometric and spectroscopic datasets will help to constrain models to unprecedented accuracy and precision, making forward modelling not only ready but also a competitive alternative to traditional methods in Stage-IV cosmology.

\acknowledgments

We thank Arne Thomsen, Alex Reeves and Pascale Berner for helpful discussions and Uwe Schmitt for informatics support.
We acknowledge the support of Euler Cluster by High Performance Computing Group from ETHZ Scientific IT Services that we used for most of our computations.

We acknowledge the use of the following software packages: 
\galsbi \citep{fischbacherGalsbiPythonPackage2025}, \texttt{numpy} \citep{vanderwaltNumPyArrayStructure2011}, \texttt{POT} \citep{flamaryPOTPythonOptimal2021}, \texttt{scikit-learn} \citep{Pedregosa:2011ork}, \texttt{pzflow} \citep{crenshawJfcrenshawPzflowV3132024,crenshawProbabilisticForwardModeling2024}, \texttt{scipy} \citep{virtanenSciPy10FundamentalAlgorithms2020}, \texttt{tensorflow} \citep{tensorflowdevelopersTensorFlow2021}, \ufig \citep{fischbacherUFigV1Ultrafast2025} and \texttt{xgboost} \citep{chenXGBoostScalableTree2016}.
Jobarrays were submitted with \texttt{esub-epipe} \citep{zurcherCosmologicalForecastNonGaussian2021,zurcherDarkEnergySurvey2022,zurcherFullCDMMapbased2023}, plots were created using \texttt{matplotlib} \citep{hunterMatplotlib2DGraphics2007} and \texttt{trianglechain} \citep{fischbacherRedshiftRequirementsCosmic2023,kacprzakDeepLSSBreakingParameter2022}.

This paper is based on data collected at the Subaru Telescope and retrieved from the HSC data archive system, which is operated by Subaru Telescope and Astronomy Data Centre (ADC) at NAOJ. 

COSMOS2020 is based on observations collected at the European Southern Observatory under ESO programme ID 179.A-2005 and on data products produced by CALET and the Cambridge Astronomy Survey Unit on behalf of the UltraVISTA consortium. 

This work has made use of data from the European Space Agency (ESA) mission {\it Gaia} (\url{https://www.cosmos.esa.int/gaia}), processed by the {\it Gaia} Data Processing and Analysis Consortium (DPAC, \url{https://www.cosmos.esa.int/web/gaia/dpac/consortium}).
Funding for the DPAC has been provided by national institutions, in particular the institutions participating in the {\it Gaia} Multilateral Agreement. 

This project used public archival data from the Dark Energy Survey (DES). Funding for the DES Projects has been provided by the U.S. Department of Energy, the U.S. National Science Foundation, the Ministry of Science and Education of Spain, the Science and Technology Facilities Council of the United Kingdom, the Higher Education Funding Council for England, the National Center for Supercomputing Applications at the University of Illinois at Urbana-Champaign, the Kavli Institute of Cosmological Physics at the University of Chicago, the Center for Cosmology and Astro-Particle Physics at the Ohio State University, the Mitchell Institute for Fundamental Physics and Astronomy at Texas A\&M University, Financiadora de Estudos e Projetos, Funda{\c c}{\~a}o Carlos Chagas Filho de Amparo {\`a} Pesquisa do Estado do Rio de Janeiro, Conselho Nacional de Desenvolvimento Cient{\'i}fico e Tecnol{\'o}gico and the Minist{\'e}rio da Ci{\^e}ncia, Tecnologia e Inova{\c c}{\~a}o, the Deutsche Forschungsgemeinschaft, and the Collaborating Institutions in the Dark Energy Survey.

The Collaborating Institutions are Argonne National Laboratory, the University of California at Santa Cruz, the University of Cambridge, Centro de Investigaciones Energ{\'e}ticas, Medioambientales y Tecnol{\'o}gicas-Madrid, the University of Chicago, University College London, the DES-Brazil Consortium, the University of Edinburgh, the Eidgen{\"o}ssische Technische Hochschule (ETH) Z{\"u}rich,  Fermi National Accelerator Laboratory, the University of Illinois at Urbana-Champaign, the Institut de Ci{\`e}ncies de l'Espai (IEEC/CSIC), the Institut de F{\'i}sica d'Altes Energies, Lawrence Berkeley National Laboratory, the Ludwig-Maximilians Universit{\"a}t M{\"u}nchen and the associated Excellence Cluster Universe, the University of Michigan, the National Optical Astronomy Observatory, the University of Nottingham, The Ohio State University, the OzDES Membership Consortium, the University of Pennsylvania, the University of Portsmouth, SLAC National Accelerator Laboratory, Stanford University, the University of Sussex, and Texas A\&M University.

Based in part on observations at Cerro Tololo Inter-American Observatory, National Optical Astronomy Observatory, which is operated by the Association of Universities for Research in Astronomy (AURA) under a cooperative agreement with the National Science Foundation.


\bibliographystyle{JHEP}
\bibliography{references}

\appendix

\section{Survey rotation}
\label{app:rotation}

For each simulated image, we prepare a host catalog with all the halos and subhalos that lie within the image boundaries.
However, we can obtain more than one independent survey footprint (except for super-survey modes) from one simulation by applying rotations.

\begin{figure}
    \centering
    \includegraphics[width=1\linewidth]{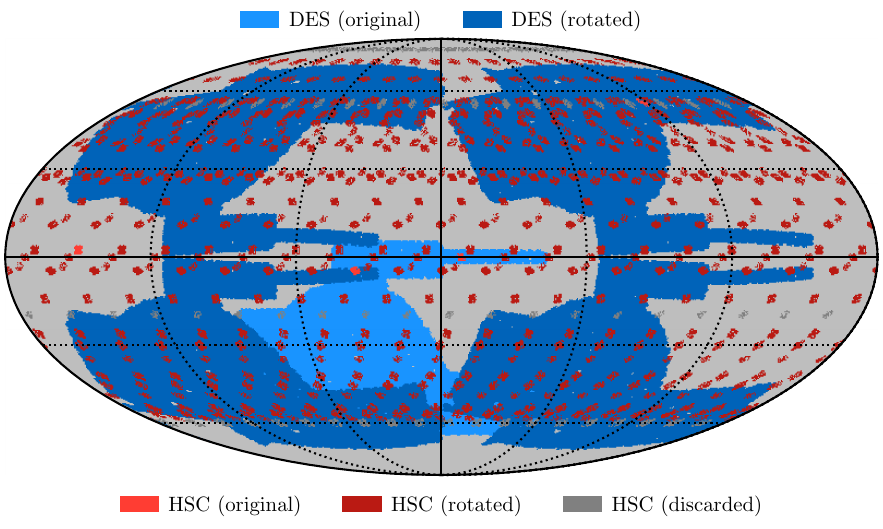}
    \caption{
    Footprints of the two surveys used in this work and their rotations to generate independent realizations of the \pinocchio simulation.
    For DES~Y3 (blue), four non-overlapping rotations are obtained following \cite{thomsenDarkEnergySurvey2026}. 
    For HSC DUD (red), $N = 160$ independent realizations are generated by translating the original footprint across the sphere on a regular grid, with discarded overlapping candidates shown in gray.
    }
    \label{fig:survey_rotation}
\end{figure}

For DES, we follow the rotation procedure from \cite{thomsenDarkEnergySurvey2026}, which creates four non-overlapping DES~Y3 footprints.
Both the original and the four rotated realizations are shown in figure \ref{fig:survey_rotation}.

The HSC DUD fields are much smaller, and we can therefore generate a large number of non-overlapping footprints.
We translate the original footprint across the sphere on a regular grid in right ascension and declination, with spacings of $\Delta\alpha = \SI{18}{deg}$ and $\Delta\delta = \SI{19}{deg}$, chosen to maximize the number of independent realizations while minimizing discarded candidates.
Candidate placements that overlap any previously accepted patch are discarded, yielding $N = 160$ independent realizations of the survey 
geometry.
The original and all rotated footprints, including discarded candidates, are shown in figure~\ref{fig:survey_rotation}.

\section{Source extraction and object selection}
\label{app:selection}
For the DES data and simulations, we use the \sextractor configuration given in 
Table~\ref{tab:sextractor}.
We then select objects from the resulting catalogs using
\begin{equation}
    \begin{aligned}
        &\mathtt{FLAGS} < 4               && \text{for } g, r, i, z, \\
        &17 < \mathtt{MAG\_AUTO} < 99     && \text{for } g, \\
        &17 < \mathtt{MAG\_AUTO} < 25.5   && \text{for } r, z, \\
        &17 < \mathtt{MAG\_AUTO} < 23.5   && \text{for } i.
    \end{aligned}
\end{equation}
Additionally, we remove objects close to bright stars and within the DES survey 
mask.
We further clean the catalogs to avoid duplicate entries from overlapping 
tile regions.

\begin{table}
  \centering
  \begin{tabular}{lc}
    \toprule
    \sextractor parameter & Value \\
    \midrule
    CATALOG\_TYPE      & FITS\_1.0 \\
    DETECT\_TYPE       & CCD \\
    DETECT\_MINAREA    & 5 \\
    THRESH\_TYPE       & RELATIVE \\
    DETECT\_THRESH     & 1.5 \\
    ANALYSIS\_THRESH   & 1.5 \\
    FILTER             & Y \\
    FILTER\_NAME       & gauss\_2.0\_5x5.conv \\
    DEBLEND\_NTHRESH   & 32 \\
    DEBLEND\_MINCONT   & 0.00001 \\
    CLEAN              & Y \\
    CLEAN\_PARAM       & 1.0 \\
    MASK\_TYPE         & CORRECT \\
    MAG\_ZEROPOINT     & 27 \\ 
    WEIGHT\_TYPE       & NONE \\
    PHOT\_APERTURES    & 12, 18 \\
    PHOT\_AUTOPARAMS   & 2.5, 3.5 \\
    PIXEL\_SCALE       & 0.168 \\
    SEEING\_FWHM       & 1.2 \\
    STARNNW\_NAME      & default.nnw \\
    BACK\_SIZE         & 128 \\
    BACK\_FILTERSIZE   & 3 \\
    BACKPHOTO\_TYPE    & LOCAL \\
    BACKPHOTO\_THICK   & 24 \\
    \bottomrule
  \end{tabular}
  \caption{\sextractor configuration used for DES~Y3 data and simulations.}
  \label{tab:sextractor}
\end{table}

Star-galaxy separation is performed using a boosted decision tree 
classifier \cite{chenXGBoostScalableTree2016} trained on image simulations using the galaxy population model from \citetalias{fischbacherGalSBIPhenomenologicalGalaxy2025} and the SHAM model from \citetalias{bernerFastForwardModelling2024}.
The input features consist of \texttt{MAG\_AUTO}, \texttt{FLUX\_RADIUS} and \texttt{CLASS\_STAR} in all bands, and the colors $g-r$, $r-i$ and $i-z$, achieving approximately 95\% accuracy, precision and recall on a balanced test set.
In the real data, where galaxies outnumber stars by orders of magnitude, stellar contamination is therefore expected to be minimal.
Crucially, since the same classifier and cuts are applied to both data and simulations, any residual contamination is consistent between the two and does not bias our inference.

For HSC, we use the same setup as in \cite{moserSimulationbasedInferenceDeep2024}; for details we refer to appendices~B and~C of this paper.

\section{Impact of the halo mass limit on position assignment}
\label{app:posassignment}
As described in section \ref{sec:sham-ot}, establishing the galaxy-halo connection and assigning realistic positions to galaxies in our image simulations requires three ingredients: a halo histogram (in our case, the full-sky halo and subhalo catalog from our simulation), a galaxy histogram (derived from the luminosity function), and a halo catalog covering the image area.
Both the halo histogram and the halo catalog are split into red and blue populations according to the model described in section \ref{sec:sham}, and are then matched to the corresponding red and blue luminosity functions separately.
The matching is performed in redshift bins of width $\Delta z = 0.01$, and the resulting OT solution is applied to the halo catalog.

However, this procedure does not guarantee a galaxy catalog that is complete down to a specified magnitude limit, because the halo histogram has a hard lower mass limit imposed by the \pinocchio simulation.
When the halo and galaxy histograms are normalized prior to matching, this mass limit translates into a faintest magnitude that can still be assigned a halo position.
Ideally, this magnitude limit is faint enough that the corresponding galaxies fall below the detection threshold in the image simulation.
However, for a realistic image simulation, undetected galaxies must also be rendered: they contribute to the sky background and to blending, both of which affect the detection and photometric measurement of brighter sources.
We therefore render galaxies down to at least one magnitude fainter than the detection limit.

\begin{figure}
\centering
\includegraphics[width=1\linewidth]{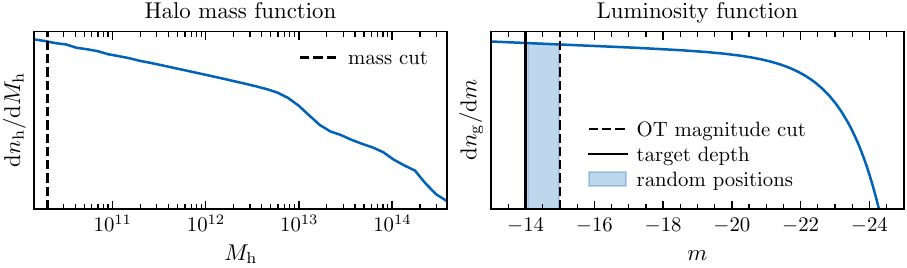}
\caption{
The \pinocchio simulation and the subhalo extraction strategy of
\citetalias{bernerFastForwardModelling2024} produce a halo mass function
with a hard lower mass limit (left panel).
Normalizing the halo and galaxy histograms prior to OT matching translates
this mass cut into a faintest magnitude that can be assigned a halo position (right panel, dashed line).
Galaxies fainter than this limit but above the target survey depth
(right panel, solid line) are assigned uniform random positions by
additionally sampling the faint end of the luminosity function;
this region is shown as the shaded area.
The values shown are chosen for illustration purposes only.
}
\label{fig:halo_lf_cuts}
\end{figure}

If, in a given redshift bin, the faintest magnitude reachable via halo-position assignment is brighter than this target limit, we additionally sample galaxies from the remaining faint end of the luminosity function down to the target magnitude and assign them uniform random positions, following the same methodology as in \citetalias{fischbacherGalSBIPhenomenologicalGalaxy2025}.
This two-step procedure is illustrated schematically in figure \ref{fig:halo_lf_cuts}: the left panel shows the halo mass function with its hard mass cut, and the right panel shows the corresponding magnitude cut in the luminosity function together with the region that is supplemented by random-position sampling.

\begin{figure}
    \centering
    \includegraphics[width=1\linewidth]{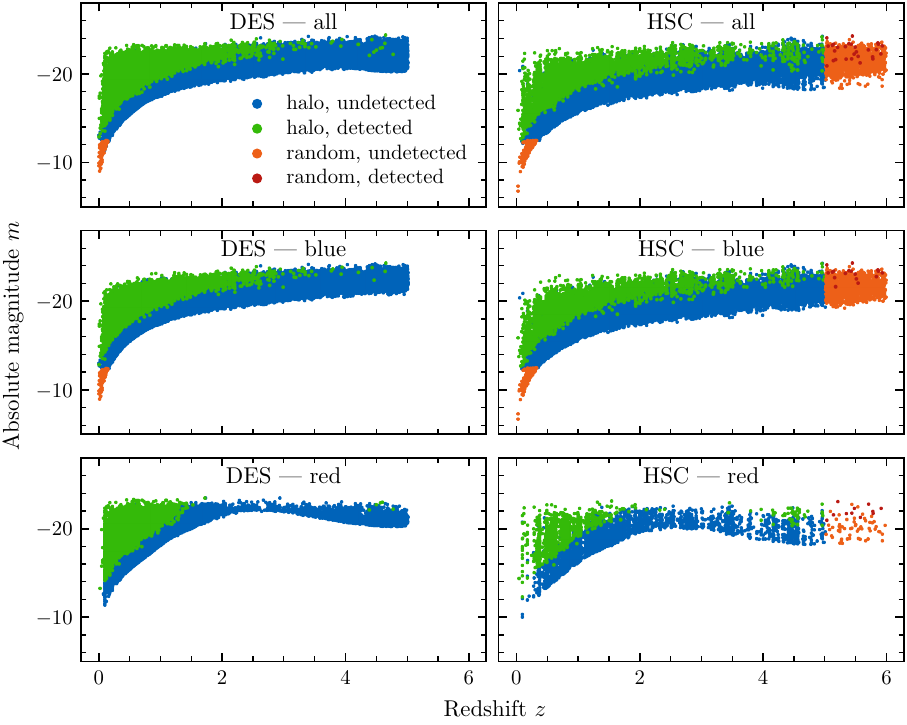}
    \caption{
        Distribution of all rendered galaxies in redshift and absolute magnitude for a DES tile (left column) and an HSC tile (right column), split into all galaxies (top), blue galaxies (middle), and red galaxies (bottom).
        Galaxies are color-coded by whether their positions were drawn from the halo catalog or assigned randomly, and whether they are detected or undetected in the image simulation.
        For DES, no detected galaxies have random position
        assignments; for HSC this fraction rises to 0.5\%.
    }
\label{fig:halo_mag_detection}
\end{figure}

In principle, this additional sampling step could be avoided entirely by using a simulation with a sufficiently low mass limit and a sufficiently high maximum redshift.
Figure \ref{fig:halo_mag_detection} shows, for both one DES and one HSC tile, the distribution of all rendered galaxies in redshift and absolute magnitude, color-coded by whether their positions were drawn from the halo catalog or assigned randomly, and whether they fall above or below the detection threshold.
In the redshift range $0.3 < z < 5$, all galaxies have halo-assigned positions.
At low redshift, even intrinsically faint galaxies are detectable, so the hard lower mass limit of the simulation translates into a relatively bright magnitude cut.
Some of our undetected galaxies at low redshift therefore have random positions.
At $z > 5$, beyond the maximum redshift of our simulation, all galaxies necessarily receive random positions.
Overall, no detected galaxies and around 0.1\% of all rendered galaxies have random position assignments for DES, where we constrain the clustering.
For HSC, 0.5\% of detected galaxies (all at $z>5$) and 4.5\% of all rendered galaxies have random position assignments.
The random position assignment therefore has a negligible impact on the clustering statistics, and all conclusions drawn in this work are robust to the mass limit of the \pinocchio simulation.

\section{Power spectrum normalization}
\label{app:cl_norm}

\begin{figure}
    \centering
    \includegraphics[width=1\linewidth]{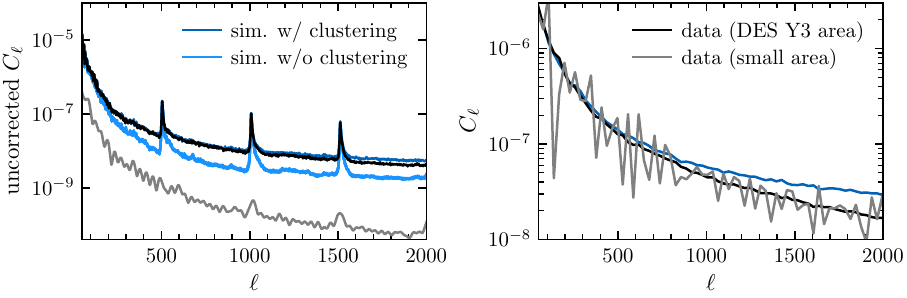}
    \caption{
    Impact of correction on power spectra.
    Left: Angular power spectrum obtained without any corrections for mask or survey depth variation for a simulation with clustering (dark blue), a simulation without clustering (light blue) and the data (black) covering the full DES Y3 footprint and the data covering a smaller area (gray). 
    Right: Angular power spectrum of data and simulation after accounting for mask, shot noise and density variations.
    The simulation without clustering has no power anymore by construction, the data power spectra measured on the smaller area is consistent with the one with larger area but with higher scatter and the simulation with clustering still has a slight excess in power at high $\ell$ as in the uncorrected case.
    Note that the simulation with clustering is not using the fiducial model of this work but one with worse agreement to better illustrate the fact that the overprediction is present in both the uncorrected and corrected case.
    }
    \label{fig:cl_norm}
\end{figure}

As described in section~\ref{sec:cl}, we correct for shot noise, masking, and survey depth variations when computing the angular power spectrum.
In figure~\ref{fig:cl_norm}, we compare power spectra for different setups. 
The data, the simulation with clustering, and the simulation without clustering all show distinct peaks at the same $\ell$ values (see figure~\ref{fig:cl_norm}, left).
These peaks originate from depth variations in the data, which we accurately forward model with our image simulation.
During inference, we simulate only a small patch of the survey, but wish to compare the resulting $C_\ell$ against the full survey data vector, which covers a larger area and is therefore less noisy.
To make this comparison valid, we apply corrections for shot noise, masking, and depth variations.
Figure~\ref{fig:cl_norm} (right) demonstrates that this correction yields consistent power spectra across different survey areas.
The overprediction of the simulation at high $\ell$, already visible in the unnormalized case, is preserved as expected.
While the simulated $C_\ell$ will remain noisier since only the small patch is generated during inference, comparing against a less noisy data vector stabilizes convergence.

\section{Redshift distributions}
\label{app:nz}
\begin{figure}
    \centering
    \includegraphics[width=1\linewidth]{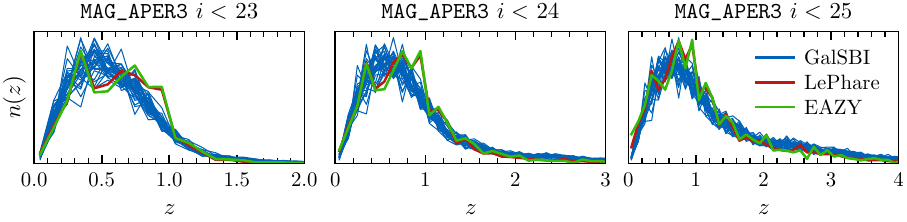}
    \caption{Same as figure \ref{fig:hsc_redshift}, but with fixed galaxy population model.}
    \label{fig:hsc_redshift_sample}
\end{figure}

To estimate the contributions of sample variance and galaxy population model uncertainty to the mean redshift, we run two simulation suites of the COSMOS field.
First, 30~simulations were generated from the same draw of the galaxy population model but with different halo catalogs.
The variation across simulations thus stems solely from the different rotations of the halo catalog; see appendix \ref{app:rotation} for details.
The corresponding redshift distributions are shown in figure \ref{fig:hsc_redshift_sample}.

\begin{figure}
    \centering
    \includegraphics[width=1\linewidth]{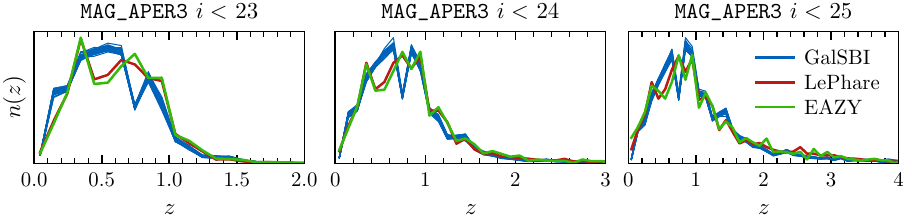}
    \caption{Same as figure \ref{fig:hsc_redshift}, but with fixed halo catalog.}
    \label{fig:hsc_redshift_model}
\end{figure}

Second, we generate 30~simulations with the same halo catalogs but marginalize over the galaxy population posterior.
These simulations therefore have the same halo positions, but the galaxy assignment and morphology modelling account for the uncertainty in the galaxy population model.
The corresponding redshift distributions are shown in figure \ref{fig:hsc_redshift_model}.

The uncertainty from the first suite accounts for sample variance of the small sky patch and would decrease when applied to a larger area.
The uncertainty from the second suite, on the other hand, reflects the uncertainty in the galaxy population model.
This uncertainty would not decrease with a larger survey area, but only if the GalSBI model can be more tightly constrained.
Assuming these two contributions are independent, we can model the total uncertainty as
\begin{equation}
    \sigma_\mathrm{total}^2 = \sigma_\mathrm{sample}^2 + \sigma_\mathrm{model}^2,
\end{equation}
where $\sigma_\mathrm{sample}$ and $\sigma_\mathrm{model}$ are estimated from the respective simulation suites.
We verify this assumption by comparing $\sqrt{\sigma^2_\mathrm{sample} + \sigma^2_\mathrm{model}}$ against the standard deviations measured from the fiducial suite in which both halo catalogs and galaxy population model draws are varied simultaneously, finding good agreement between the two standard deviations.
\section{Constraints on galaxy-halo connection parameters}
\label{sec:galpop_gh}

\begin{figure}
    \centering
    \includegraphics[width=\linewidth]{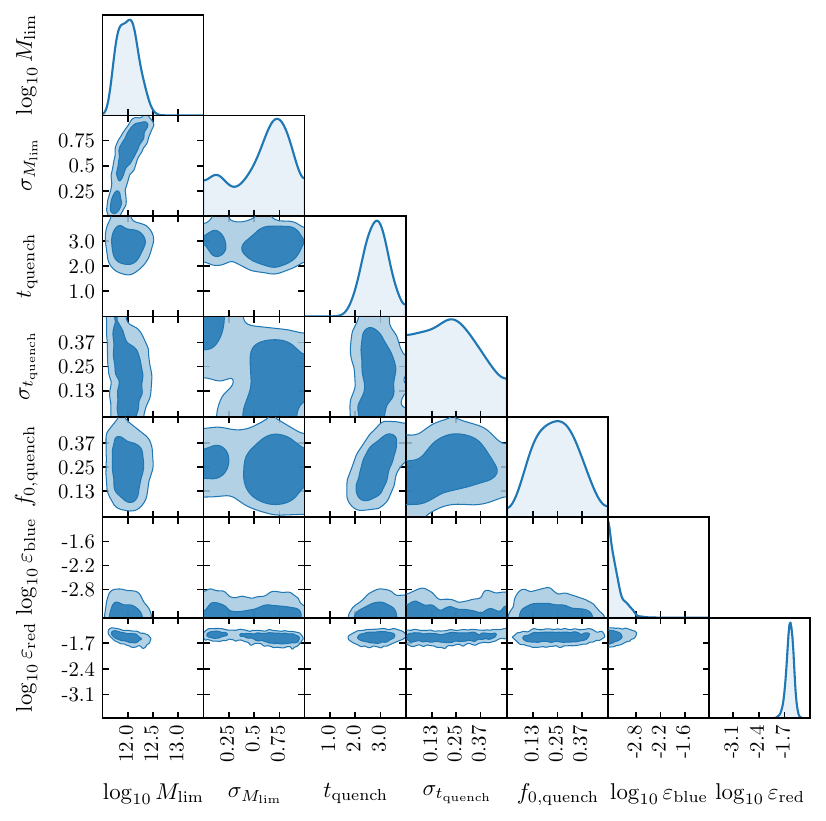}
    \caption{
    Constraints on the parameters of the galaxy-halo connection.
    The plotting limits correspond to the prior range.
    }
    \label{fig:galpop_gh}
\end{figure}

In figure \ref{fig:galpop_gh}, we show the constraints on the parameters of the galaxy-halo connection.
Other galaxy population parameters are consistent with the constraints from \citetalias{fischbacherGalSBIPhenomenologicalGalaxy2025} and therefore not shown here.

\end{document}